\PassOptionsToPackage{pdfpagelabels=false,pdftex}{hyperref} 
\documentclass[fleqn,usenatbib]{mnras}
\usepackage[T1]{fontenc}
\usepackage{ae,aecompl}
\usepackage{graphicx}	
\usepackage{amsmath}	
\usepackage{amssymb}	

\usepackage{times}
\usepackage{color}
\usepackage[caption=false]{subfig}
\usepackage{xspace}
\usepackage{algorithm}
\usepackage{multirow}

\definecolor{customhdrcolor}{rgb}{0.0,0.0,0.0}
\definecolor{customcitecolor}{rgb}{0.0,0.5,0.75}
\definecolor{customlinkcolor}{rgb}{0.0,0.5,0.75}

\hypersetup{colorlinks=true,linkcolor=customlinkcolor,urlcolor=customlinkcolor,citecolor=customcitecolor}

\ifpdf\else\usepackage{graphics}\fi

\newcommand{\degree}{\ensuremath{^{\circ}}\xspace}

\DeclareRobustCommand{\TUSSEN}[3]{#2}

\title[An optimized algorithm for multi-scale wideband deconvolution]{An optimized algorithm for multi-scale wideband deconvolution of radio astronomical images}

\def\ASTRON{$^{1}$}
\def\Rhodes{$^{2}$}
\def\SKASA{$^{3}$}
\author[A.~R.~Offringa \& O. Smirnov]{A.~R.~Offringa\ASTRON\thanks{Corresponding author. E-mail: offringa@astron.nl},
O. Smirnov\Rhodes$^,$\SKASA
\\
\ASTRON{}Netherlands Institute for Radio Astronomy (ASTRON), PO Box 2, 7990 AA Dwingeloo, The Netherlands\\
\Rhodes{}Department of Physics and Electronics, Rhodes University, PO Box 94, Grahamstown, 6140, South Africa\\
\SKASA{}SKA South Africa, 3rd Floor, The Park, Park Road, Pinelands, 7405, South Africa
}

\begin{document}

\date{Accepted 2017 June 19. Received 2017 June 16; in original form 2017 March 16.}
\pagerange{\pageref{firstpage}--\pageref{lastpage}}
\pubyear{2016}

\label{firstpage}
\maketitle

\begin{abstract}
We describe a new multi-scale deconvolution algorithm that can also be used in multi-frequency mode. The algorithm only affects the minor clean loop. In single-frequency mode, the minor loop of our improved multi-scale algorithm is over an order of magnitude faster than the \textsc{casa} multi-scale algorithm, and produces results of similar quality. For multi-frequency deconvolution, a technique named joined-channel cleaning is used. In this mode, the minor loop of our algorithm is 2-3 orders of magnitude faster than \textsc{casa msmfs}. We extend the multi-scale mode with automated scale-dependent masking, which allows structures to be cleaned below the noise. We describe a new scale-bias function for use in multi-scale cleaning. We test a second deconvolution method that is a variant of the \textsc{moresane} deconvolution technique, and uses a convex optimisation technique with isotropic undecimated wavelets as dictionary. On simple, well calibrated data the convex optimisation algorithm produces visually more representative models. On complex or imperfect data, the convex optimisation algorithm has stability issues.
\end{abstract}


\section{Introduction}
Radio astronomical interferometers sample Fourier modes of the sky. By calculating the inverse Fourier transform of the interferometric visibilities, a sky map can be made. Because an interferometric array samples the Fourier plane incompletely, sky images are convolved with the point spread function (PSF) of the array. To obtain accurate and sensitive flux density measurements of the sky, deconvolution is required.

The most commonly used deconvolution methods in radio astronomy are variants of H\"ogbom Clean \citep{hogbom-clean}. Variants of H\"ogbom Clean have been developed that deconvolve structures with different scales \citep{wakker-1988-multires-clean,multiscale-clean-cornwell-2008}, take frequency information into account \citep{sault-1994-mf-deconvolution} or both \citep{rau-msmfs-2011}. These variants use the basic principle of H\"ogbom Clean, i.e., they search for the highest peak and subtract its contribution iteratively.

H\"ogbom Clean is a relatively fast algorithm, and thanks to optimizations such as Clark Clean \citep{clark-clean}, deconvolution has so far not been a major bottleneck in the imaging of interferometric radio observations. However, modern observatories such as LOFAR \citep{lofar-2013} and the expanded Karl G. Jansky Very Large Array (VLA) \citep{expanded-jvla-2011} have high spatial resolution, large band-widths and large field of views, which necessitates the use of slower, more complex algorithms such as multi-scale clean \citep{multiscale-clean-cornwell-2008} and \textsc{msmfs} \citep{rau-msmfs-2011} on large image sizes, and need to satisfy higher dynamic-range requirements. This increase of requirements, in combination with the design of faster gridders \citep{offringa-wsclean-2014}, has made it important to improve the performance of deconvolution algorithms.

The recent emergence of the field of compressed sensing has provided alternative deconvolution approaches that are fundamentally different from Clean \citep{wiaux-2009-compr-sens}. Several deconvolution schemes have been developed that are based on compressed sensing ideas, such as Purify \citep{carrillo-2014-purify}, \textsc{moresane} \citep{dabbech-2015-moresane} and \textsc{sasir} \citep{girard-2015-sasir}. Another recent development is the introduction of Bayesian deconvolution methods \citep{junklewitz-2016-resolve}.

So far, these new methods have only been validated with simple test scenarios, where the volume of the data is kept small and some of the practical issues are ignored. The involved practical issues are however limiting the use of these methods in realistic scenarios. Such issues are: i) Many current convex optimization implementations call the visibility-sky forward and backward functions tens to hundreds of times, which is infeasible for large data sets (without significant parallelization and distribution efforts); ii) the so-called $w$-term \citep{perley-noncoplanar-arrays} or beam and ionospheric effects \citep{aprojection-2008} that break the assumption of a spatially invariant PSF are not taken into account; iii) perfectly calibrated data are used. However, at low frequencies or high dynamic ranges, calibration errors, ionospheric disturbances and beam modelling errors are inevitable; and iv) spectral variation is not taken into account during deconvolution. It has been shown that this can be taken into account using more complex deconvolution schemes \citep{ferrari-2015-mf-regularization, junklewitz-2015-mfs}, but thus far only for relatively simple test cases.

In this paper, we present a fast multi-scale algorithm that can be extended to perform fast multi-frequency deconvolution, and focus on the applicability of it on real, imperfect data. We extend the multi-scale multi-frequency algorithm to include an automatic scale-dependent masking technique for reaching a higher dynamic range. Furthermore, we compare results with the \textsc{moresane} sparse optimization technique, and present an extension to \textsc{moresane} that takes spectral variation into account. We start by describing the new algorithms in Sec.~\ref{sec:methods}, present results in Sec.~\ref{sec:results} and discuss the results and draw conclusions in Sec.~\ref{sec:conclusions}.

\section{Methods}\label{sec:methods}
In this section, we describe our new multi-scale and multi-frequency deconvolution algorithms. All described algorithms except the \textsc{moresane} algorithm are implemented in the C++ language and integrated in \textsc{wsclean}, which is available under an open-source license\footnote{\textsc{wsclean} and its manual can be found at \url{http://wsclean.sourceforge.net/}.}. The \textsc{wsclean} imager uses a $w$-stacking algorithm for the gridding, which is particularly advantageous for wide-field imaging \citep{offringa-wsclean-2014}.

\subsection{Multi-frequency deconvolution in \textsc{wsclean}} \label{sec:joined-channel-deconvolution}
When the required dynamic range of imaging is high, source-intrinsic spectral variation and a spectrally-varying primary-beam response can make it necessary to account for spectral variation in the deconvolution.
A commonly used method to take spectral information into account during cleaning, is by using a generalization of the Sault-Wieringa method \citep{sault-1994-mf-deconvolution}. This method is used in the \textsc{casa} \texttt{clean} task when \texttt{nterms} is set higher than one. For each requested Taylor-expansion term, inversions are performed to construct dirty and PSF images. During cleaning, the spectral variation is fitted and the dirty image is corrected for the spectral variation.

\textsc{wsclean} uses a different approach to multi-frequency deconvolution. The method is somewhat similar to the multi-frequency deconvolution approach  in Obit \citep{cotton-2008-obit}. In our approach, the full bandwidth is divided into output channels. These are normally formed by dividing the bandwidth into evenly-spaced subbands, but such spacing is not necessary for the method to work. The algorithm starts by making a dirty and PSF image for each output channel, by imaging the input channels that fall within the frequency range of the output channel. An image-based clean loop is now started with a few alterations: i) the component to be cleaned is determined by finding the highest peak in the integrated image; ii) the spectral variation of this component is determined by measuring the flux in the individual channel images at the location of the peak; and iii) before subtracting the PSFs from the dirty images, a function can be fitted to the spectral measurements. Typical functions to be fitted are a polynomial to enforce smoothness, or a sinusoid to fit the rotation measure of a polarized source. The fit is performed at each step of the minor cycle, and the flux subtracted from the image at each 
output channel is given by the fitted value. The fitting step is optional, and can be left out, in which case any spectral shape is allowed.

This multi-frequency deconvolution method is called ``joined-channel deconvolution'' in \textsc{wsclean}. Essentially, the approach uses the full bandwidth to determine the location of clean components, which decreases the chance of selecting noise peaks or sidelobes. In the joined-channel method, separate model images are constructed for each output channel, and stopping thresholds are relative to the integrated image. The Cotton-Schwab method \citep{cotton-schwab-clean} can be used in the normal way to correct the $w$-terms and/or other instrumental direction-dependent behaviour.

For inversion and prediction, the channelized representation of the data can be converted to polynomial functions. This allows interpolation of the channelized data, either by performing inversion and prediction with a higher number of channels or by inverting and predicting the coefficients with the appropriate Taylor-expansion factors, as described by \citet{sault-1994-mf-deconvolution} and \citet{rau-msmfs-2011}. To perform either method of interpolation, a spectral polynomial needs to be found that describes the intrinsic flux of the radio source. When the polynomial is of second order or higher, and the channel bandwidth is not negligibly small, it is not sufficient to fit a polynomial over the bandwidth-integrated values. Essentially, this is because the channel average is not equal to the channel value at the channel-central frequency. Instead, a channel integrated flux density $S_i$ for channel $i$ that integrates over frequencies from $\nu_i$ to $\xi_i$ is given by
\begin{small}\begin{eqnarray}
\notag
 S_i & = & \frac{1}{\xi_i-\nu_i}\int\limits_{\nu=\nu_i}^{\xi_i} c_1 + c_2 \nu + c_3 \nu^2 + ... \\
\label{eq:integrated-polynomial}
 & = & \sum\limits_{t=1}^{N_\textrm{terms}} c_t \frac{\xi_i^t - \nu_i^t}{t \left( \xi_i - \nu_i \right) }
\end{eqnarray}\end{small}%
Estimating the $N_\textrm{terms}$ unaveraged polynomial coefficients $c_t$ with $1 \le t \le N_\textrm{terms}$ given the $M$ bandwidth-integrated flux measurements $S_i$ with $1 \le i \le M$ and $M\ge N_\textrm{terms}$ requires the least-squares minimalization of a system of $M$ equations given by Eq.~\ref{eq:integrated-polynomial}, instead of ordinary polynomial regression. The combination of source-intrinsic and primary beam spectral curvature can be fitted by this function. By correcting each channel image independently for the primary beam before fitting, source-intrinsic spectra can be obtained.

\subsection{Multi-scale clean in \textsc{wsclean}} \label{sec:wsclean-multiscale}
\citet{multiscale-clean-cornwell-2008} presented a multi-scale clean approach that is now commonly used. Multi-scale clean is considered an important improvement of the standard H\"ogbom clean algorithm, as it decreases the problem of negative bowls around bright resolved structures and has better convergence properties \citep{rich-2008-multiscale-things}.

\begin{algorithm}
\caption{Fast multi-scale algorithm in pseudocode}
\label{multiscale-algorithm}
\textbf{function} minorIterationLoop($\mathcal{R}$, $\mathcal{P}$) \\
\textit{Input:}\\
\hspace*{3mm} Residual image $\mathcal{R}$ \\
\hspace*{3mm} PSF image $\mathcal{P}$ \\
\textit{Using:}\\
\hspace*{3mm} Set of scales $\mathcal{A}$ \\
\hspace*{3mm} Scale-kernel function $K_\alpha(\mathbf{x})$ for all scales $\alpha$ \\
\hspace*{3mm} Scale-bias function $S(\alpha)$ \\
\textit{Output:}\\
\hspace*{3mm} Model image $\mathcal{M}$ \\

\textbf{while} no stopping criterion reached \textbf{do}\\
\{        \\
\hspace*{3mm} \textbf{foreach} scale $\alpha \in \mathcal{A}$ \\
\hspace*{3mm} \{           \\
\hspace*{6mm} $\mathcal{R}_\alpha \leftarrow \mathcal{R} \ast K_\alpha$ \textit{\hspace*{5mm}\{scale-convolved residual\}}\\
\hspace*{6mm} $p_\alpha \leftarrow \max \mathcal{R}_\alpha$ \textit{\hspace*{5mm}\{peak value of $\mathcal{R}_\alpha$ \}}\\
\hspace*{3mm} \}                                      \\
\hspace*{3mm} Find the most significant scale $\tilde{\alpha} = $argmax$_\alpha \left( p_\alpha S(\alpha) \right)$ \\
\hspace*{3mm} Execute the subminor loop with: \\
\hspace*{4mm} - $\mathcal{R}_{\tilde{\alpha}}$ as residual image \\
\hspace*{4mm} - $\mathcal{P} \ast K_{\tilde{\alpha}} \ast K_{\tilde{\alpha}}$ as PSF \\
\hspace*{4mm} - $M'$, an image to be filled with new components \\
\\
\hspace*{3mm} $\mathcal{M} \leftarrow \mathcal{M} + M' \ast K_{\tilde{\alpha}}$ \\
\hspace*{3mm} $\mathcal{R} \leftarrow \mathcal{R} - \mathcal{P} \ast M' \ast K_{\tilde{\alpha}}$ \\
\} \\
\end{algorithm}

\begin{algorithm}
\caption{Multi-scale subminor loop in pseudocode}
\label{multiscale-subminor-loop}
\textbf{function} subMinorLoop($\mathcal{R}$, $\mathcal{P}$) \\
\textit{Input:}\\
\hspace*{3mm} Residual image $\mathcal{R}$ \\
\hspace*{3mm} PSF image $\mathcal{P}$ \\
\textit{Using:}\\
\hspace*{3mm} Scale-bias function $S(\tilde{\alpha})$ \\
\hspace*{3mm} Multi-scale loop gain $g_1$ \\
\hspace*{3mm} Clean gain $g_2$ \\
\textit{Output:}\\
\hspace*{3mm} Model image $M$ \\

$\tilde{v} \leftarrow v \leftarrow \max \mathcal{R}$ \textit{\hspace*{5mm}\{peak value of $\mathcal{R}$ \}}\\
Let $A$ be the set of all pixels with value $v'$: $(1-g_1)v \leq v' \leq v$ \\
\textbf{while} $\tilde{v} \geq (1-g_1) v$ $\wedge$ no stopping criterion reached \textbf{do} \{        \\
\hspace*{3mm} $\tilde{v} \leftarrow$ largest value in $A$ \\
\hspace*{3mm} $p \leftarrow$ position of $\tilde{v}$ \\ 
\hspace*{3mm} Make $\mathcal{P}'$ by translating $\mathcal{P}$ to $p$ and multiplying it by $\tilde{v}g_2$ \\
\hspace*{3mm} Subtract $\mathcal{P}'$ from all pixels in $A$  \\
\hspace*{3mm} $M[p] \leftarrow M[p] + \tilde{v}g_2$ \textit{\hspace*{5mm}\{Add model component at position $p$\}} \\
 \} \\
\end{algorithm}

In this section, we will describe a new, fast multi-scale algorithm. This algorithm has been implemented in \textsc{wsclean}. A pseudocode description is given in Algorithm~\ref{multiscale-algorithm}. The multi-scale algorithm replaces the normal H\"ogbom image-based cleaning step, while the Cotton-Schwab major iterations are performed as usual. In our algorithm, the minor loop is further divided into subminor iterations, and during these subminor iterations, only one particular scale is considered. The algorithm starts by making dirty images at the considered scales by convolving the dirty input image with the corresponding scale kernels. The scale to be cleaned is selected by finding the convolved image with the highest maximum. Once the scale has been selected, the PSF is convolved with the selected scale and a subminor loop is started that performs a number of clean iterations with the current scale.

Once the maximum peak in the convolved dirty image for that scale has been reduced by a certain gain value\footnote{This is the \texttt{-multiscale-gain} parameter in \textsc{wsclean}} (typically 10\%--20\%), the subminor loop finishes and the algorithm reconsiders which scale to clean until a major-iteration cleaning criterion has been reached. Stopping criteria include the major loop gain, the number of iterations, the cleaning threshold, or finding a negative component.

The subminor loop iterations are essentially normal H\"ogbom clean iterations: the dirty image is searched for a peak, a component is added to the model image and the PSF is subtracted with a specified gain from the dirty image. The difference between a H\"ogbom clean loop and the subminor loop, is that in the subminor loop, the input image is an image convolved with a scale, and the PSF is convolved with the scale twice. The latter is because the response to a structure of a given scale in the input image is the PSF convolved with that scale, and the response in an image convolved with that scale is therefore convolved twice. Once the subminor loop finishes, the computed model components are convolved with the scale kernel and added to the full model.

Because the subminor loop corresponds to a normal H\"ogbom clean iteration, an optimization similar to the Clark optimization \citep{clark-clean} can be used to decrease the computational cost of the algorithm. A pseudocode implementation for an optimized subminor loop is given in Algorithm~\ref{multiscale-subminor-loop}. In our implementation, the subminor loop selects all dirty image pixels with amplitude between the peak and the stopping level of the loop. Then, in each iteration, the maximum of the selected pixels is found, the corresponding PSF pixels times the gain is subtracted from the selected pixels, and the components are added to the image. When the subminor loop finishes, the model components are convolved with the PSF and subtracted from the full input image.

\begin{figure}
\begin{center}
\includegraphics[width=7.8cm]{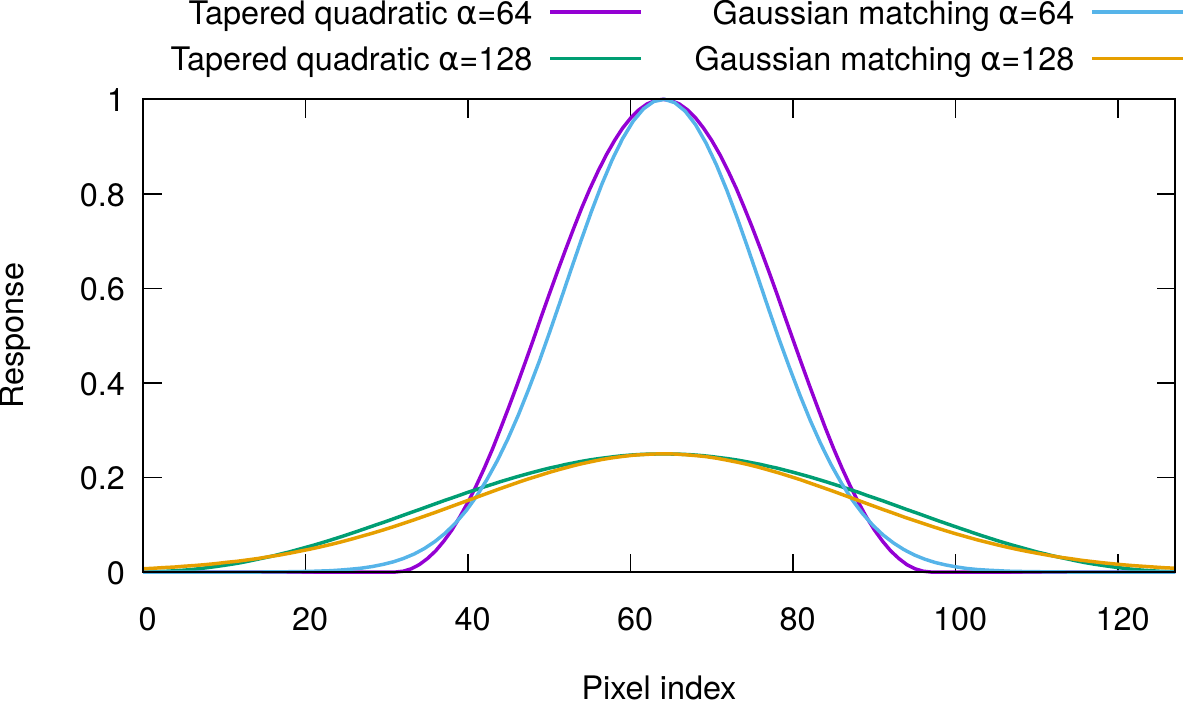}
\caption{Shape functions for scales $\alpha=64$ pixels and $\alpha=128$ pixels.}
\label{fig:multiscale-shape-function}
\end{center}
\end{figure}

We have tested two different functions as the scale convolution kernel. The first function is the tapered quadratic function as defined by \citet{multiscale-clean-cornwell-2008}:
\begin{equation}
K_\alpha(\mathbf{x}) = \frac{1}{\mu_\alpha}
H(\frac{\|\mathbf{x}\|}{\alpha}) \left[1.0 - \left(\frac{\|\mathbf{x}\|}{\alpha}\right)^2 \right].
\end{equation}
Here, $\mathbf{x}$ is the two-dimensional position in the kernel, $H$ is a Hann window function and $\alpha$ is the scale size in pixels. $\mu_\alpha$ is a normalization term, such that the integral of the kernel equals unity. The advantage of this shape function is that it has a limited support, and the convolutions therefore require fewer evaluations of the function. The second shape function that we have implemented is the Gaussian function. For the Gaussian width parameter $\sigma$ a value of $\sigma=3/16\alpha$ is used, which approximately matches the width of the tapered quadratic function. This implies that the full-width half-maximum (FWHM) of the Gaussian is $2\sqrt{2 \ln 2}(3/16)\alpha\approx0.45\alpha$. The advantage of the Gaussian function is that the Fourier transform can be analytically calculated, and Gaussian shapes are therefore often supported as primitive shapes in sky models. Examples for both functions with $\alpha=64$ and $\alpha=128$ are plotted in Fig.~\ref{fig:multiscale-shape-function}.

The scale sizes that are used in the cleaning are normally automatically calculated from the $uv$-coverage. Alternatively, the user can provide a list of scales. The automated calculation works as follows: As a first scale, the zero-scale $\alpha_0$ (delta-scale) is always used. The next scale, $\alpha_1$ is selected to have a full-width window size of four times the smallest scale in the image, which is calculated from the longest baseline. The full-width window size of this scale is therefore four times the FWHM of the synthesized beam size. The factor of four was empirically determined to be a reasonable value, and implies that the FWHM (in contrast to its window size) of the first shape $\approx 1.8$ times the FWHM of the synthesized beam. When scales smaller than this size are added, some point sources are cleaned with this scale instead of the delta scale. Finally, the scale sizes following $\alpha_1$ are the double of its predecessor, $\alpha_{i+1}=2\alpha_i$, and scales are added until the scale function no longer fits inside the image.

It is straightforward to extend the described multi-scale algorithm to use spectral information, by using the joined-channel approach described in \ref{sec:joined-channel-deconvolution}. In the minor and subminor iterations, the joined-channel multi-scale approach decides which scale and location to clean using the full bandwidth, and measures the strength of the scale in each individual frequency channel.

\subsubsection{Scale-bias function}
\citet{multiscale-clean-cornwell-2008} has introduced a scale-bias function that balances the preference between the selection of large and small scales. \citet{rau-msmfs-2011} mention this bias function is an empirical approximation of the inverse volume of the scale kernel, which normalizes the response to a given scale. They describe a method that calculates the response of a specific scale without using the scale-bias function. In the \textsc{moresane} approach \citep{dabbech-2015-moresane}, a scale is selected based on the significance of the peak, by calculating its signal-to-noise ratio. We take a different approach, and let a scale-bias function influence two aspects of the algorithm: i) when to deconvolve a scale; and ii) how the global clean threshold stopping criterion specified for the delta scale is propagated to the larger scales. In deciding \textit{when} to deconvolve a scale, a bias function needs to make sure that the smallest fitting scale is selected for a source of a particular size and that the large scales are cleaned before they cause incorrect small-scale detections. This balancing is in particular required because the kernel functions for different scales are not orthogonal. In deciding the stopping criterion for a scale, the scale bias should avoid cleaning a scale when it is not significant. This is particularly important because radio interferometers often have different sensitivities at different scales. While the selection preference and stopping criterion can be decoupled by using different bias functions for each, they are closely related. Because of the non-orthogonal scales, it is not desirable to clean a significant scale whilst less sensitive scales are still present at a much higher level, because these will influence the selection of significant scales as well. Therefore, for \textsc{wsclean} we have chosen to use a single bias function. We note that astronomers will often balance the sensitivity per scale by applying appropriate image weighting \citep{briggs-1995}, which can also be used as a mechanism to optimize multi-scale cleaning. An imaging weighting scheme that properly selects the scales of interest and at the same time balances the sensitivity between the scales, helps the cleaning and avoids having to optimize the scale stopping criterion bias or scale selection preference bias.

Because our algorithm uses a large range of scales by default, the scale-bias function $S(\alpha) = 1-0.6\alpha/\alpha_\textrm{max}$ described by \citet{multiscale-clean-cornwell-2008} does not work very well. If two deconvolutions are performed with this bias function that have different maximum scales but are otherwise identical, the relative preference between two small scales is different between the two runs. To avoid this, we modify the scale bias function and keep the bias-ratio between two consecutive scales constant:
\begin{equation} \label{eq:scale-bias}
 S(\alpha_i) = \begin{cases}
 1 & \text{ if }\alpha_i=0 \\
 \beta^{-1-\log_2 \alpha_i / \alpha_1} & \text{ otherwise.}  
\end{cases}
\end{equation}
with $\beta$ the scale bias level\footnote{This is the \texttt{-multiscale-scale-bias} parameter in \textsc{wsclean}.}. Lower values will clean larger scales earlier and deeper. For the set of scales $[0, 8, 16, 32, 64]$, this formula will produce corresponding biases of $[1, \beta, \beta^2, \beta^3, \beta^4]$. The default value for $\beta$ is 0.6, which we found to work well on a wide range of observations. This choice is further discussed in \S\ref{sec:scale-bias-results}. 

\subsubsection{Scale-dependent subtraction gain}
During each subminor iteration, a scale kernel is deconvolved from the dirty image by subtracting the scale-convolved PSF multiplied by the component value and the gain from the dirty image and adding the scale kernel to the model, again multiplied by the component value and gain. If the subtraction gain for different scales is set to the same value, the algorithm spends a relatively small number of iterations on the large scales because of the larger integrated flux value of the large scale kernel. To increase the large-scale accuracy of the algorithm, a per-scale gain is used that is lower for larger scales. While it might seem appropriate to scale the gain to the inverse of the kernel volume, doing so results in spending too many subminor iterations on the large scales. In \textsc{wsclean}, the scale gain $g_\alpha$ is set to the volume of the scale kernel divided by the peak of the scale-convolved PSF, normalized such that $g_0$, the gain for scale zero, equals the user-requested gain (which defaults to 0.1) :
\begin{equation}
 g_\alpha = g_0 \frac{\mu_\alpha \left( \textrm{PSF} \otimes {K_0} \right) [0, 0]}{\mu_0 \left( \textrm{PSF} \otimes {K_\alpha} \right) [0, 0]}.
\end{equation}
The indexing operation $[0, 0]$ represents selection of the central pixel and $\mu$ is defined as before. The reasoning behind this formula is that, were only a single scale to be used, the number of iterations required for that scale would be approximately equal to the number of iterations required for cleaning a resized image with delta scales, if the resized image had a pixel size corresponding to that particular scale.

\subsubsection{Automatic scale-dependent masking} \label{sec:automated-masking-description}
One of the problems of the Clean algorithm is that it leaves residuals behind with a flux density level equal to the stopping threshold. Because of this, the flux of sources is systematically underestimated in the recovered model. When this model is used for self-calibration, sources lose flux (which is absorbed in the gain solutions), causing what is known as ``self-calibration bias''. There are two commonly employed solutions for this. The first method is to run a source detector on the restored images, and use the fitted sources as the self-calibration model. Since the restored images include the residuals, a source detector takes any undeconvolved flux into account. One issue with this method, is that it commonly fails for fields with diffuse emission (e.g. the Galactic plane), which is not properly detected and/or represented by source detectors. The second method is to use a mask, and decrease the threshold. This can typically lower the threshold from $\sim 3\sigma$ to $0.3-0.5 \sigma$, thereby decreasing the bias substantially. Typically, masks are created by hand or with the help of a source detector. One approach is to image the data without a mask, then make the mask using the initial image and finally rerun the imaging with this mask and a deeper threshold. This is the method currently employed in the Factor pipeline \citep{vanweeren-factor-2016} for direction-dependent calibration of LOFAR data.

Here we present a third option: we have implemented a deconvolution algorithm that initially cleans down to a first threshold and simultaneously makes a mask, and then continue down to a second, deeper threshold using the constructed mask. For single-scale H\"ogbom Clean, the implementation of this is trivial: once the first threshold is reached, a mask is created containing all currently found components. In the second phase, only components are cleaned that have already been detected. For multi-scale cleaning, automatically creating a mask is slightly more complicated. If the same method was used, large components that are found during cleaning would be put into the mask in their entirety, thereby providing almost no constraints on cleaning of smaller scales within those large scales, causing overfitting of the small scales. Therefore, in our implementation a mask is created for each scale. When a larger scale is cleaned, the centre position of the scale is marked in the mask for that scale. Once the first threshold has been reached and the mask is used, the algorithm is only allowed to place components of a particular scale at a position at which that scale was already cleaned. Intuitively, this process can be considered to fit existing components to their residuals. In the first phase, it places components into the model that are significantly above the noise, and in the second phase it makes sure the components represent the data accurately.

A further extension that we implemented in \textsc{wsclean} is to clean relative to a local sliding-window RMS value. Because calibration artefacts increase the local RMS, this helps in preventing the selection of these artefacts as components. Once the first threshold is reached and the scale-dependent mask is applied, cleaning is no longer relative to the local RMS, so that sources in high-RMS areas are still modelled with most of their flux density. Use of a local RMS is important for self-calibration of low-frequency observations, which can exhibit very strong ionospheric artefacts around bright sources. Simultaneously, due to the large field-of-view of low-frequency observations, other parts of the field of view may contain fainter sources that are still strong enough to require deconvolution. Local RMS thresholding is in particular important for the LOFAR low-band antenna (LBA) survey (De Gasperin, in prep.).

\subsection{\textsc{wsclean} with \textsc{moresane} deconvolution}
\textsc{moresane} \citep{dabbech-2015-moresane} is a convex optimisation algorithm that uses the isotropic undecimated wavelet transform (\textsc{iuwt}; \citealt{strarck-2007-iuwt}) to benefit from sparsity in the transformed space. \textsc{moresane} has an iterative approach. During an iteration, it selects pixels with significant signal, uses a conjugate gradient solver to deconvolve these pixels and subsequently subtracts a part of the obtained model convolved by the PSF from the dirty image. The latter uses a subtraction gain similar to the Clean algorithm. A Python implementation of the \textsc{moresane} algorithm is available\footnote{Written by J. Kenyon, available at \url{https://github.com/ratt-ru/PyMORESANE}} \citep{kenyon-2015-pymoresane}. The Python implementation works entirely in image space and assumes the PSF is position-independent.

Because of $w$-terms, $A$-terms and gridding effects, a PSF is often not position independent. To extend the \textsc{moresane} implementation to support a position-dependent PSF, we have build \textsc{wsclean} around \textsc{moresane} with a Cotton-Schwab approach. This approach starts by making initial dirty and PSF images. Then, \textsc{moresane} is executed up to some threshold. The resulting model is forward-modelled by running a prediction-imaging sequence. Subsequently, the initial model is convolved with the PSF and added to the residual image, and \textsc{moresane} is executed again with a deeper threshold. This is repeated several times.

\textsc{moresane} does not support multi-frequency deconvolution. The only way to take frequency information into account during deconvolution, is to image the observation at different frequencies, and run \textsc{moresane} on each frequency. A system was implemented in \textsc{wsclean} to fit a spectral polynomial or logarithmic polynomial to the resulting models and use this during the prediction. This is done each time the prediction is performed. While this mitigates spectral curvature, it does not allow cleaning as deeply as the joined-channel deconvolution (\S\ref{sec:joined-channel-deconvolution}) or the multi-term deconvolution strategies, because the deconvolution does not use the full bandwidth to determine the significant pixels.

\begin{figure*}
\begin{center}
\begin{tabular}{ccc}
 \subfloat[\textsc{casa} multi-scale]{\includegraphics[width=7cm]{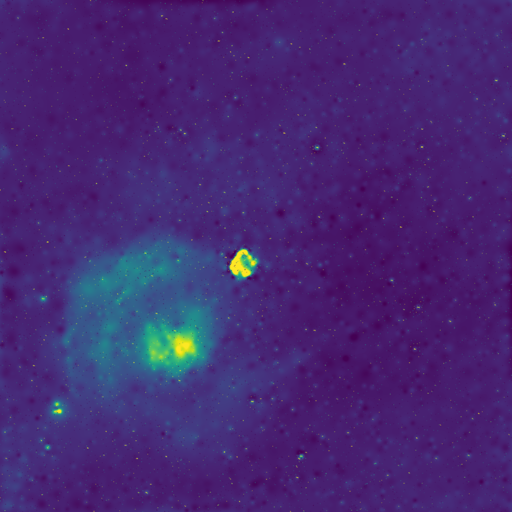}} &
 \subfloat[\textsc{wsclean} multi-scale ($\beta$=0.60)]{\includegraphics[width=7cm]{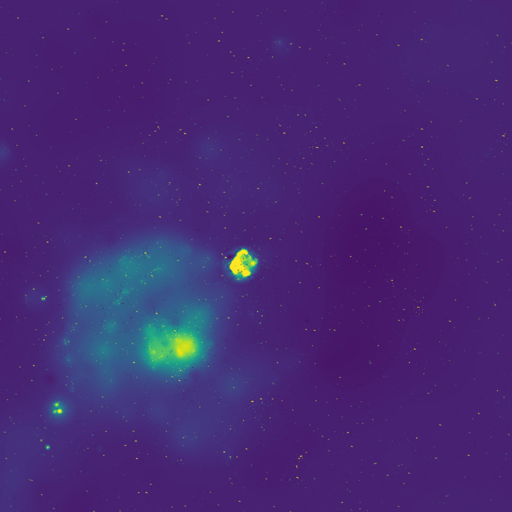}} &
 \multirow{-2}{*}[4cm]{\includegraphics[width=1.3cm]{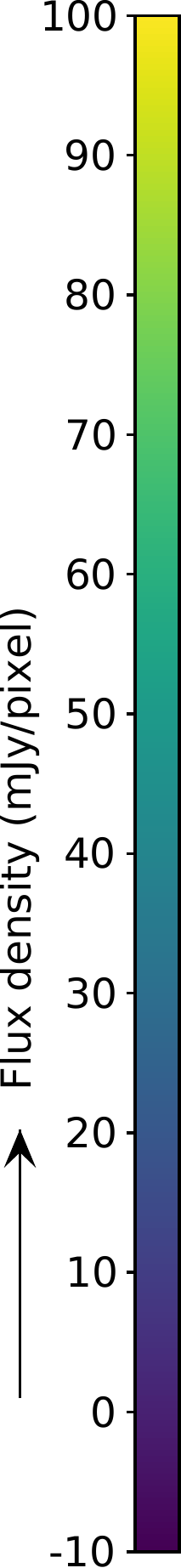}} \\
 \subfloat[\textsc{wsclean} \textsc{iuwt}]{\includegraphics[width=7cm]{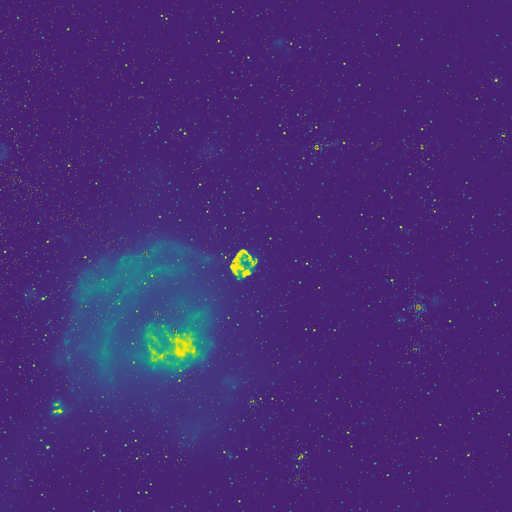}} &
 \subfloat[\textsc{wsclean} + \textsc{moresane}]{\includegraphics[width=7cm]{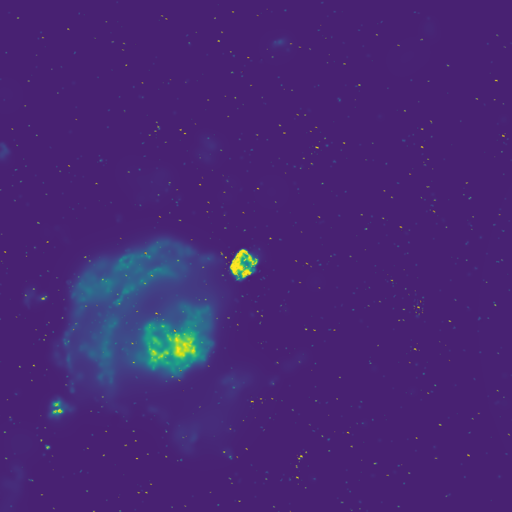}} &
 \\
\end{tabular}
\end{center}
\caption{Model images showing 17 $\times$ 17\degree of an MWA observation of supernova remnants Vela and Puppis A, displayed with the same colour scale. Despite the 3$\sigma$ threshold, noise components appear in the \textsc{casa}, \textsc{wsclean} and \textsc{iuwt} models. \textsc{wsclean} has a stronger preference for the delta-function scale, while \textsc{casa} cleans many components with a slightly larger scale. The \textsc{iuwt} and \textsc{moresane} models model the structures of the supernova remnants with sharper details.}
\label{fig:vela-model-images}
\end{figure*}

\begin{figure*}
\begin{center}
\begin{tabular}{ccc}
 \subfloat[\textsc{casa} (rms=64~mJy/PSF)]{\includegraphics[width=7cm]{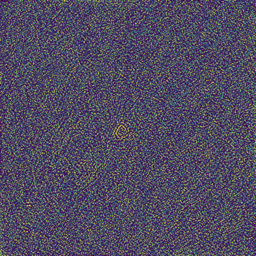}} &
 \subfloat[\textsc{wsclean} multi-scale with $\beta$=0.60 (rms=50~mJy/PSF)]{\includegraphics[width=7cm]{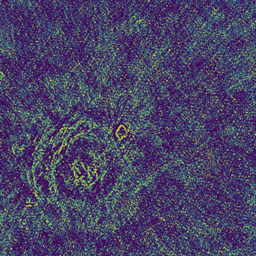}} &
 \multirow{-2}{*}[6cm]{\includegraphics[width=1.2cm]{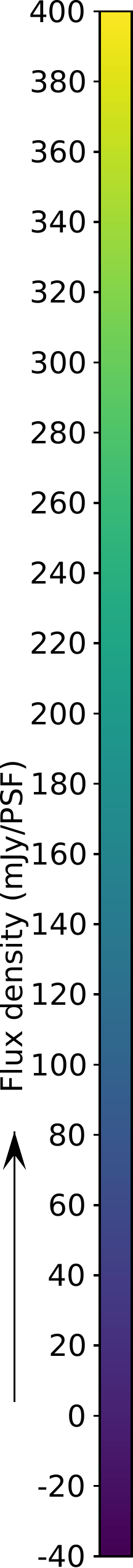}} \\
 \subfloat[\textsc{wsclean} \textsc{iuwt} ($\sigma$=63~mJy/PSF)]{\includegraphics[width=7cm]{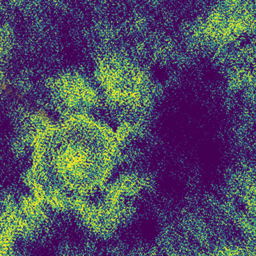}} &
 \subfloat[\textsc{wsclean} + \textsc{moresane} ($\sigma$=75~mJy/PSF)]{\includegraphics[width=7cm]{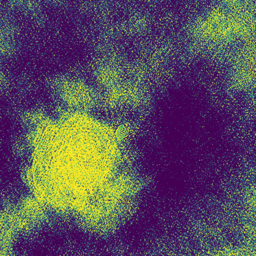}} &
 \\
\end{tabular}
\end{center}
\caption{Residual images for the MWA observation of supernova remnants Vela and Puppis A, displayed with the same colour scale.}
\label{fig:vela-residual-images}
\end{figure*}

\begin{figure*}
\begin{center}
 \subfloat[\textsc{wsclean} multi-scale model ($\beta$=0.35)]{\includegraphics[height=7cm]{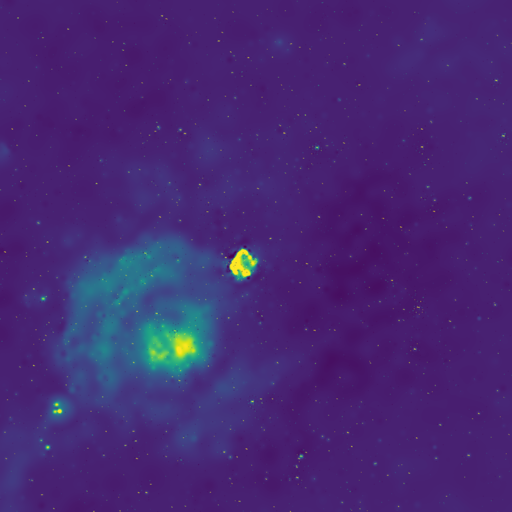}}\
 \includegraphics[height=7cm]{img/axes/vela-model-axis}\hspace{5mm}
 \subfloat[\textsc{wsclean} multi-scale residual ($\beta$=0.35, rms=62~mJy/PSF)]{\includegraphics[height=7cm]{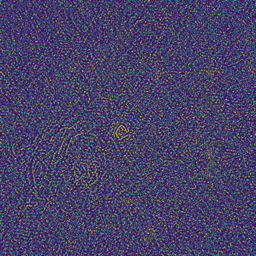}}\
 \includegraphics[height=7cm]{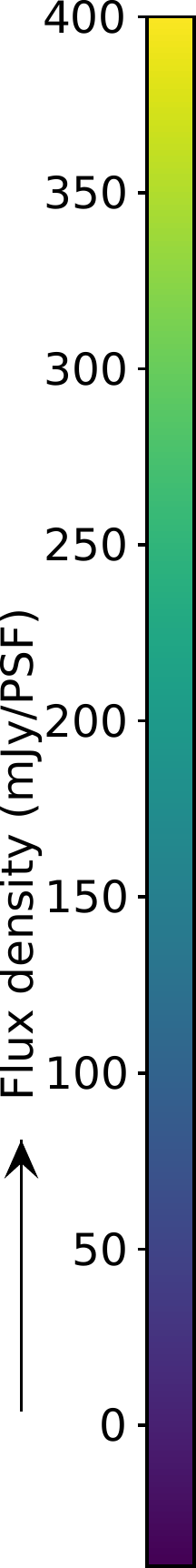}
\end{center}
\caption{Result of the \textsc{wsclean} multi-scale method with a multi-scale bias of $\beta=0.35$, showing that \textsc{wsclean} produces a residual image that more closely resembles the \textsc{casa} results of Figs.~\ref{fig:vela-model-images} and \ref{fig:vela-residual-images} with these settings. To make this result, the largest two scales have been turned off, as these cause the algorithm to diverge due to the strong bias for the large scales. While these images show less residual structure than the $\beta=0.60$ results (see Fig.~\ref{fig:vela-residual-images}), the larger scales are cleaned into the noise and cause the algorithm to become unstable.}
\label{fig:vela-casa-vs-wsclean}
\end{figure*}

\begin{figure*}
\begin{center}
 \subfloat[\textsc{wsclean} multi-frequency, multi-scale with $\beta$=0.6 (rms=1.4~mJy/PSF)]{\includegraphics[height=7cm]{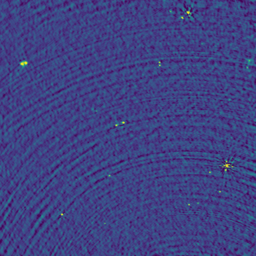}}\
 \includegraphics[height=7cm]{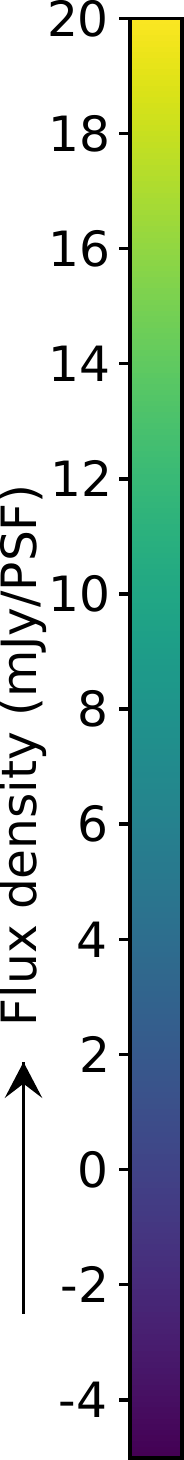}\hspace{5mm}
 \subfloat[\textsc{wsclean} multi-frequency, \textsc{iuwt} (rms=2,7~mJy/PSF)]{\includegraphics[height=7cm]{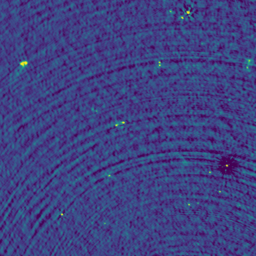}}\
 \includegraphics[height=7cm]{img/axes/iuwt-residual-axis}
\end{center}
\caption{Restored images of part of the LOFAR EoR 3c196 field at high resolution (1.5'' pixel size, 10'' synthesized beam). The source to the right and below the vertical centre has a strong apparent brightness of 1~Jy. This makes the ionospheric perturbations and primary beam changes relevant when cleaning this source down to the noise. The image centre is r.a. $8^\circ23'00''$, dec. $48^\circ02'00''$.}
\label{fig:lofar-blob}
\end{figure*}

\subsection{Multi-frequency deconvolution with \textsc{iuwt} convex optimisation}
We have written an extended multi-frequency implementation of the \textsc{moresane} methodology, that allows multi-frequency deconvolution in a manner similar to joined channel deconvolution described in \S\ref{sec:joined-channel-deconvolution}. The algorithm works similar to \textsc{moresane}: it is iterative, and each iteration starts by selecting pixels and scales based on their significance. After this, a convex optimisation is performed using the conjugate gradient method to deconvolve these pixels. These steps are performed on the full-bandwidth images, so that selection and deconvolution is performed with the full sensitivity and uv-coverage of the observation. To accommodate the model that is found in an iteration to the spectral variations, the model is segmented into connected components, and each connected component is fit to each frequency. The fitted connected components are added to the model for each frequency and its convolved contribution is subtracted from the residual images.

The aim of this method is to deconvolve fields that contain mostly point-sources, but might have a few resolved structures, and include spectral information. Modelling of Epoch of Reionization fields is a typical use-case, as target fields for EoR experiment are chosen to have few resolved sources \citep{ncp-eor-yatawatta,offringa-mwa-deep-eor-survey}. At the same time, spectral information is essential for these experiments. For large, complex sources that require different spectral indices at different positions, this method will not suffice, since it will estimate only one spectral index for each connected structure. 

\begin{figure*}
\begin{center}
 \subfloat[Original]{\includegraphics[height=4.8cm]{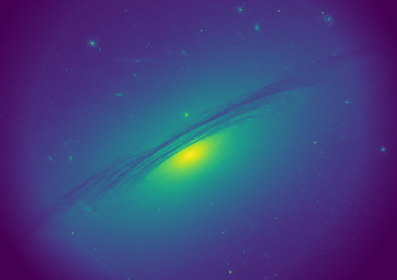}}\,
 \includegraphics[height=4.8cm]{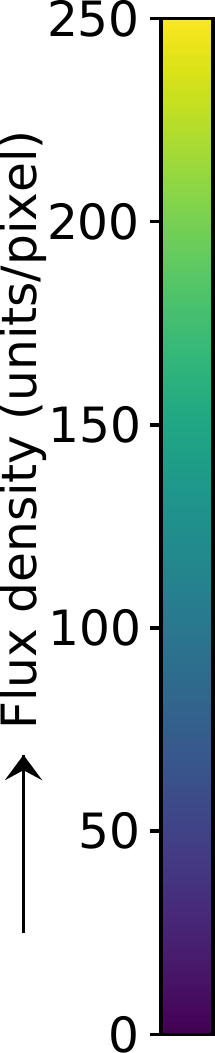}\hspace{5mm}
 \subfloat[Convolved image ($\sigma$=640,000~units/PSF)]{\includegraphics[width=6.9cm]{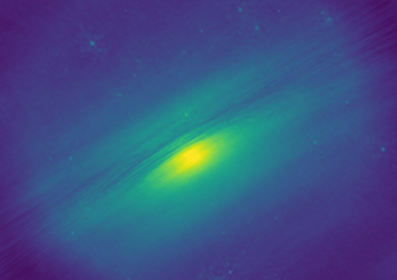}}\,
 \includegraphics[height=4.8cm]{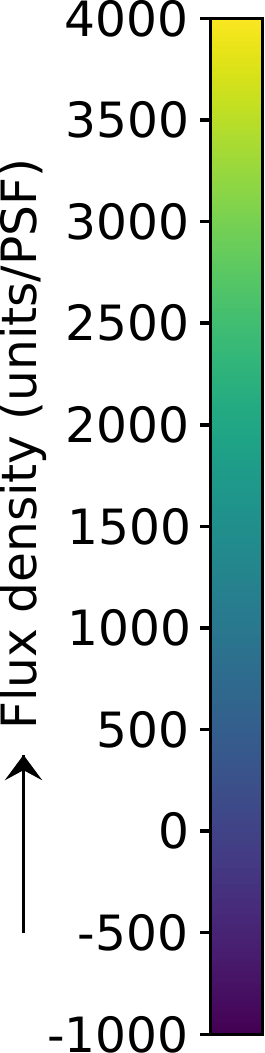}\\
 \subfloat[\textsc{casa} model]{\includegraphics[width=6.9cm]{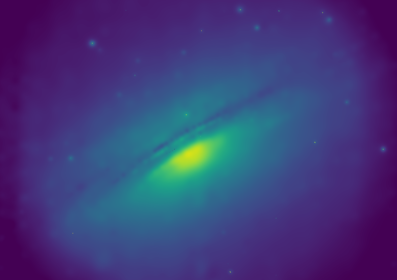}}\,
 \includegraphics[height=4.8cm]{img/axes/ugc12591-model-axis}\hspace{5mm}
 \subfloat[\textsc{casa} residual ($\sigma$=37~units/PSF)]{\includegraphics[width=6.9cm]{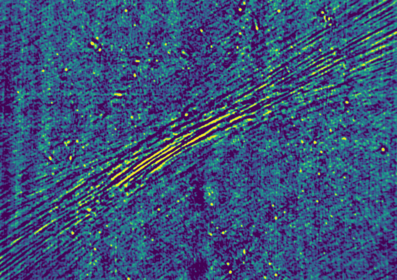}}\,
 \includegraphics[height=4.8cm]{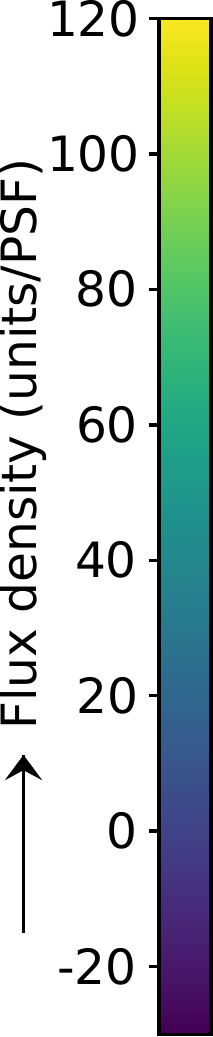}\\
 \subfloat[\textsc{wsclean} model]{\includegraphics[width=6.9cm]{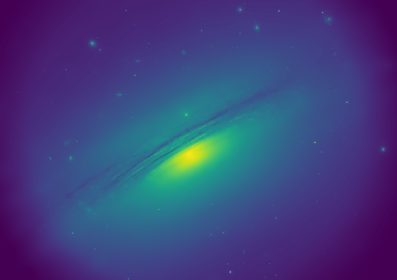}}\,
 \includegraphics[height=4.8cm]{img/axes/ugc12591-model-axis}\hspace{5mm}
 \subfloat[\textsc{wsclean} residual ($\sigma$=15~units/PSF)]{\includegraphics[width=6.9cm]{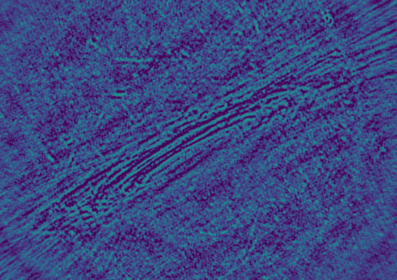}}\,
 \includegraphics[height=4.8cm]{img/axes/ugc12591-residual-axis}\\
 \subfloat[\textsc{wsclean iuwt} model]{\includegraphics[width=6.9cm]{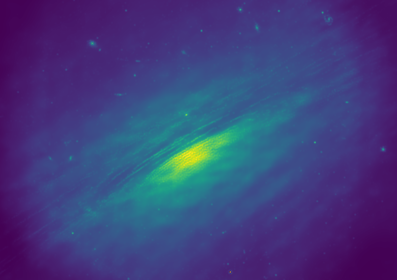}}\,
 \includegraphics[height=4.8cm]{img/axes/ugc12591-model-axis}\hspace{5mm}
 \subfloat[\textsc{wsclean iuwt} residual ($\sigma$=19~units/PSF)]{\includegraphics[width=6.9cm]{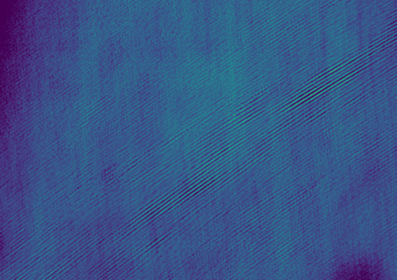}}\,
 \includegraphics[height=4.8cm]{img/axes/ugc12591-residual-axis}\\
\end{center}
\caption{Deconvolution comparison using a ground truth input image. For this comparison we used a simulation of the MWA observing a scaled version of UGC12591. The original image is an adapted ESA/Hubble \& NASA image.}\label{fig:ugc12591-comparison}
\end{figure*}

\section{Results} \label{sec:results}
In this section, we will test the methods described in the previous section on real and simulated data.

\subsection{\textsc{casa} and \textsc{wsclean} multi-scale clean comparison} \label{sec:casa-wsclean-comparison}
As was described in \S\ref{sec:wsclean-multiscale}, the multi-scale clean algorithm in \textsc{wsclean} uses a different approach compared to the algorithm described by \citet{multiscale-clean-cornwell-2008}: In our algorithm, the deconvolved scale is kept constant inside the subminor iterations. In this section we check whether this optimization produces different results.

Fig.~\ref{fig:vela-model-images} shows model images for a 2-min observation of the Vela and PupA supernova remnants recorded with the Murchison Widefield Array (MWA), made with different deconvolution strategies. The residual images are shown in Fig.~\ref{fig:vela-residual-images}. In these figures, the first and second images were made with the multi-scale clean implementation of \textsc{casa} and \textsc{wsclean}, respectively. Both images were made with default settings.

There are some differences between the \textsc{casa} and \textsc{wsclean} deconvolutions: in the model images, it is apparent that \textsc{casa} has cleaned a particular scale into the noise. In the residual image, the \textsc{casa} result shows some small-scale residuals, but no large-scale residual structure. The residual structure in \textsc{wsclean} is more evident, both at small and large scales. While it is generally desirable to have no residuals after deconvolution, a property of Clean is that when a proper (bias-compensated) threshold is used, residuals will be visible. Consequently, in this case \textsc{casa} has cleaned specific scales of the image into the noise. This is also likely the cause of the slightly higher RMS of the image (64 mJy/PSF in \textsc{casa} versus 50 mJy/PSF in \textsc{wsclean}).

While the different results between \textsc{casa} and \textsc{wsclean} could indicate that the \textsc{wsclean} constant-scale subminor-loop optimization produces different results, it is more likely a result of different stopping criteria caused by the scale bias. To analyze this, we have tried to tweak the scale bias to make the \textsc{wsclean} model more similar to the \textsc{casa} model. Fig.~\ref{fig:vela-casa-vs-wsclean} shows the results for $\beta=0.35$, which biases the deconvolution to clean large scales earlier and deeper compared to the previous results for $\beta=0.60$. With the scale bias of 0.35, \textsc{wsclean} does not converge. To be able to make the algorithm converge, we remove the two largest scales. The resulting residual image of \textsc{wsclean} is more similar to the residual image of \textsc{casa}. In particular, no residual large-scale structure is visible. Quantitatively, \textsc{wsclean} produces a slightly lower residual RMS with both $\beta=0.35$ (47 mJy/PSF) and $\beta=0.60$ (50 mJy/PSF) compared to \textsc{casa} with default settings (64 mJy/PSF), but it is likely that the small-scale bias of \textsc{casa} can be tweaked to produce a lower residual RMS. The difference between \textsc{casa} and \textsc{wsclean} that is present even after tweaking $\beta$ to produce more similar results is most likely because \textsc{wsclean} uses Eq.~\ref{eq:scale-bias} to calculate the scale bias in \textsc{wsclean}, rather than the scale-bias function given by \citet{multiscale-clean-cornwell-2008}. The two imagers also use different gridders, which can create slightly different results \citep{offringa-wsclean-2014}. Because the results of the new multi-scale algorithm are quantitatively not worse, it can be concluded that the constant-scale subminor-loop optimization has no negative effect.

\subsection{Multi-scale, \textsc{iuwt} and \textsc{moresane} comparison}
The bottom two images of Figs.~\ref{fig:vela-model-images} and \ref{fig:vela-residual-images} show the model and the residual images of the Vela/Puppis A field using the \textsc{iuwt} and \textsc{moresane} methods. Compared to the multi-scale results, the model images look sharper and contain more detail of, for example, the ripples inside the supernova remnant. The \textsc{iuwt} model image contains many false single-pixel components, while \textsc{moresane} does not. This is most likely because of a different scale selection and stopping criteria, since otherwise the methods implement the same underlying algorithm. The details of the model  of the supernova remnants in \textsc{iuwt} are mostly similar to the \textsc{moresane} results. The most apparent features in the residual images are large-scale structures that have not been cleaned. The \textsc{iuwt} residual image with an RMS of 63~mJy/PSF has approximately the same RMS compared to \textsc{casa}'s multi-scale clean residual, while \textsc{moresane} leaves a higher residual RMS of 75~mJy/PSF. The model and residual images show that the \textsc{moresane} based methods are better at capturing the sharp small-scale variations of the sources, but do not lead to a deeper image.

While using \textsc{iuwt} and \textsc{moresane} on real, imperfect observations, we have encountered cases in which these methods do not converge properly. To demonstrate such a case, we apply the methods to a LOFAR observation at 150~MHz and image the observation at a resolution of 1.5''. At this resolution, direction-dependent effects from the ionosphere cause sharp spoke-like artefacts around bright sources. Fig.~\ref{fig:lofar-blob} shows the (multi-frequency) results for multi-scale and \textsc{iuwt}. This test-case shows that multi-scale clean is robust to calibration errors: while it does not improve the image around bright sources, it also does not diverge, and is able to clean other fainter sources in the field correctly. The \textsc{iuwt} algorithm however diverges on the strong source with calibration errors. Because of this, the algorithm is not able to clean other sources sufficiently. Running \textsc{moresane} on the image results in the same problem. We have seen that this is a consistent problem with the \textsc{moresane}-based approaches, which makes it difficult to use \textsc{moresane} and \textsc{iuwt} on high-resolution LOFAR data or inside a self-calibration loop that needs to deal with imperfect data. The diverging pattern seen in Fig.~\ref{fig:lofar-blob} (right image) reflects the positivity constraint imposed by \textsc{iuwt}. In fact, a similar divergence occurs when a positivity constraint is applied during H\"ogbom or multi-scale cleaning. Both \textsc{iuwt} and \textsc{moresane} allow the positivity constraint to be turned off, but this makes the algorithm diverge faster.

To compare the methods using an observation for which the ground truth is known, we simulated an MWA observation of a scaled version of UGC12591. While this is not a realistic observable for the MWA, the image consists of diffuse structure, sharp features as well as point sources, and is therefore an excellent test image that highlights the differences between the methods. To simulate the observation, we applied a primary beam to the HST image and predicted it into an MWA observation.

The results for the UGC12591 test set are shown in Fig.~\ref{fig:ugc12591-comparison}. \textsc{moresane} does not converge on this simulation, and its results are left out. The \textsc{casa} and \textsc{wsclean} multi-scale results are made with the same clean stopping thresholds and the default scale-bias parameters. \textsc{casa} shows a preference for cleaning with larger scales, with a model image that misses sharp features. It also shows wave-like features in the background of the model and residual, which is likely due to overcleaning of the large scales. \textsc{wsclean} captures more detail in the model and produces lower residuals. It more often uses point components. As before, these differences are likely to be the result of the different scale-bias functions that are used in \textsc{casa} and \textsc{wsclean}, and are not a fundamental difference between the different algorithms. Tweaking the \textsc{casa} scale bias parameter might improve the result. The \textsc{wsclean iuwt} residuals are reasonably low and without structure, but the \textsc{wsclean iuwt} model shows clear wave-like features which are not present in the original image.

\begin{figure}
\begin{center}
\includegraphics[width=8.5cm]{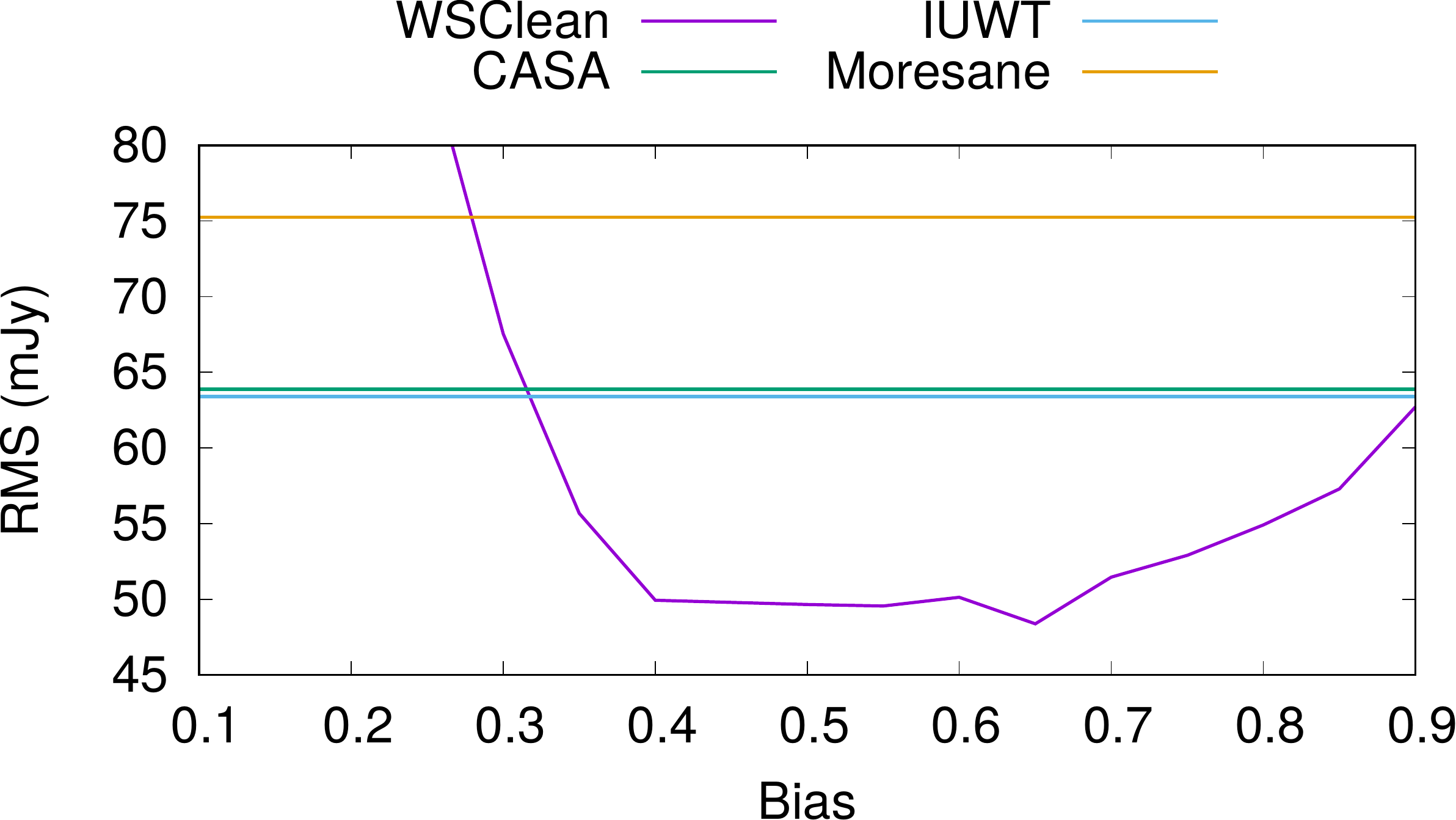}
\caption{Residual RMS values after multi-scale cleaning of the Vela/Puppis~A field with various values for the bias levels $\beta$. The default value for $\beta$ in \textsc{wsclean} is $0.6$.}
\label{fig:multiscale-bias-vs-rms}
\end{center}
\end{figure}

\subsection{Scale bias} \label{sec:scale-bias-results}

\begin{figure*}
\begin{center}
\subfloat[Multi-scale model image without masking]{\includegraphics[width=7cm]{img/wsclean-bias0_60-model}}\,
\includegraphics[height=7cm]{img/axes/vela-model-axis}\hspace{5mm}
\subfloat[Multi-scale model image with automatic masking]{\includegraphics[width=7cm]{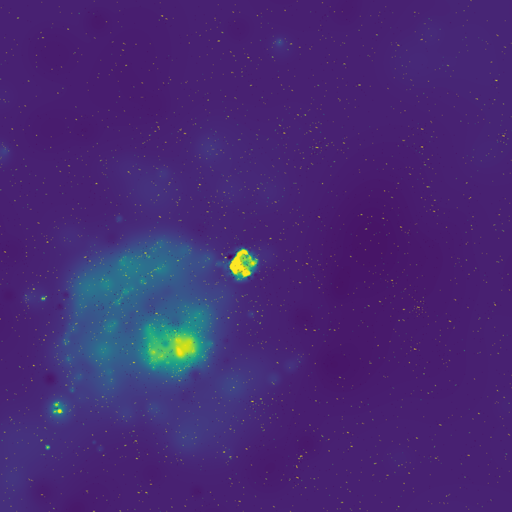}}\,
\includegraphics[height=7cm]{img/axes/vela-model-axis}\\
\subfloat[Multi-scale residual without masking (rms=50~mJy/PSF)]{\includegraphics[width=7cm]{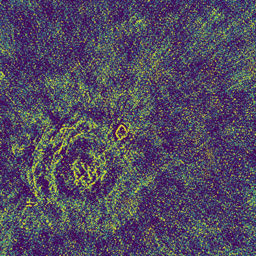}}\,
\includegraphics[height=7cm]{img/axes/vela-residual-axis-small}\hspace{5mm}
\subfloat[Multi-scale residual with automatic masking (rms=38~mJy/PSF)]{\includegraphics[width=7cm]{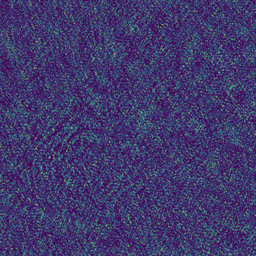}}\,
\includegraphics[height=7cm]{img/axes/vela-residual-axis-small}
\end{center}
\caption{Comparison between normal multi-scale clean and multi-scale clean with automatic scale-dependent masking.}
\label{fig:auto-mask}
\end{figure*}
\begin{figure*}
\begin{center}
\subfloat[Reconstructed model image]{\includegraphics[height=5cm]{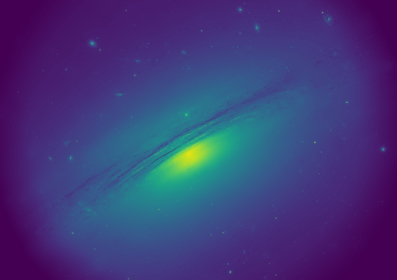}}\,
\includegraphics[height=5cm]{img/axes/ugc12591-model-axis}\hspace{5mm}
\subfloat[Residual image ($\sigma$=8.6~units/PSF)]{\includegraphics[height=5cm]{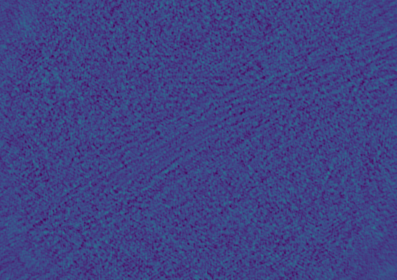}}\,
\includegraphics[height=5cm]{img/axes/ugc12591-residual-axis}
\end{center}
\caption{Automatic scale-dependent masking applied on the UGC12591 test-set.}
\label{fig:auto-mask-ugc12591}
\end{figure*}

Eq.~\ref{eq:scale-bias} describes the scale bias function used by \textsc{wsclean}. As was shown in \S\ref{sec:casa-wsclean-comparison}, the clean bias of the multi-scale algorithm has a significant effect on the deconvolution. When the bias is not properly chosen, certain scales can be over-cleaned. Fig.~\ref{fig:vela-casa-vs-wsclean} demonstrates this. We will now analyze some further properties of the scale-bias function used by \textsc{wsclean}.

Evaluating Eq.~\ref{eq:scale-bias} with scales $\alpha_A$ and $\alpha_B$ such that $\alpha_B=2\alpha_A$ results in $S(\alpha_B) = \beta S(\alpha_A)$. Hence, when each subsequent scale is twice the size of its preceding scale (as is the default for \textsc{wsclean}), the scale-bias parameter $\beta$ defines the bias factor between two subsequent scales. Furthermore, the convolution with a scale kernel is essentially a matched filter, and when a feature in the image exactly matches the scale $\alpha_A$, the result will be a peak response of
\begin{equation}\label{eq:response-small-kernel}
r_A = \int_{\textbf{x} \in N} K_A^2(\textbf{x}),
\end{equation}
with $K_A$ the scale kernel for scale $\alpha_A$ and $N$ the set of pixels in the area of the kernel. The next larger scale kernel $\alpha_B$ will match less well with this feature, and results in a peak response of 
\begin{equation}\label{eq:response-larger-kernel}
r_B = \int_{\textbf{x} \in N} K_A(\textbf{x}) K_B(\textbf{x}).
\end{equation}
Now, when $r_A \le \beta r_B$, this feature would be cleaned by a scale of $A$ or smaller, and when $r_A > \beta r_B$ it will be cleaned by scale $B$.
Therefore, the ratio between Eqs.~\eqref{eq:response-small-kernel} and \eqref{eq:response-larger-kernel} is the scale bias boundary at which the algorithm starts selecting large scales when encountering a feature which is in fact smaller. This ratio can be calculated by simple evaluation of the formula, and is approximately 0.5 for the smallest scales and converges to 0.38 for the large scales. This sets an approximate limit on the scale bias: it should be higher than 0.5, because a lower bias would cause small scales to be cleaned with large kernels.

Similarly, if a feature exactly matches the larger scale $\alpha_B$, it will result in a response of $\int_{\textbf{x}} K_B^2(\textbf{x})$ to kernel $B$ and a response of $\int_{\textbf{x}} K_A(\textbf{x}) K_B(\textbf{x})$ to kernel $A$. The ratio between these two formulae is approximately 0.7 at small scales and converges to 0.66 at large scales. Since such a feature should be cleaned with scale $\alpha_B$, the bias parameter $\beta$ should not be set higher than 0.66.

In practice, selecting a bias value outside these boundaries can still result in cleaning with both large and small kernels, because the PSF introduces negative values. If these negative values are in the response area of kernel $K_B$, but not in $K_A$, they will lower the value of $r_B$ without affecting $r_A$. Nevertheless, these calculations do provide a range of reasonable bias levels. The fact that these boundaries remain approximately constant independent of the selected scale sizes is an important argument to favour the use of the scale bias function of Eq.~\eqref{eq:scale-bias} over the function defined by \citet{multiscale-clean-cornwell-2008}.

We have cleaned the Vela/Puppis~A set with various values for the bias parameter $\beta$ and measured the RMS of the residual image. The results are given in Fig.~\ref{fig:multiscale-bias-vs-rms}. Indeed, the range $0.5 < \beta < 0.66$ provides the best results in terms of the RMS, although the RMS for values of $\beta$ between $0.35$--$0.5$ or $0.66$--$0.75$ is only slightly higher. For \textsc{wsclean}, we have chosen a default value of $\beta=0.6$. Empirically this value works well, and we have not come across an observation that requires a different value. Since the weighting also effects the relative strength between large and small scales, using different weightings can be used to further tweak the scale bias of multi-scale clean.

\subsection{Results of scale-dependent masking}
The automated scale-dependent masking algorithm as described in \S\ref{sec:automated-masking-description} was tested on the Vela/Puppis~A set. Cleaning without a mask to 0.3$\sigma$ is not possible for this observation, as it causes the clean to diverge at approximately $2\sigma$. The left image of Fig.~\ref{fig:auto-mask} shows the result for cleaning up to 3$\sigma$ without mask, while the right image shows the results after continuing this clean to 0.3$\sigma$ with the automatically-created scale-dependent mask. Cleaning with an automated mask converges well. The unmasked imaging shows structure from Vela and Puppis A, while the residual image of the masked algorithm is much more noise-like. The masking decreases the RMS of the residual image from 50~mJy/PSF to 38~mJy/PSF. The contribution of system noise, as measured from the Stokes~V image, is estimated to be 36~mJy for this observation.

Fig.~\ref{fig:auto-mask-ugc12591} shows the reconstructed and residual images for the UGC12591 test set using scale-dependent masking. These can be compared with the results in Fig.~\ref{fig:ugc12591-comparison}. The model image shows a close reconstruction of the original image, although visually not much different from the unmasked \textsc{wsclean} multi-scale result. The masked residual image shows low noise-like residuals with almost no structure.

\begin{figure*}
 \subfloat[Single-scale clean (residual RMS=880 $\mu$Jy/PSF)]{\includegraphics[width=5cm]{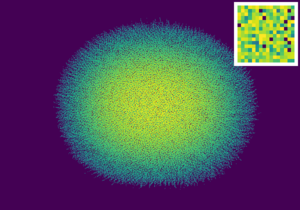}\,\includegraphics[width=5cm]{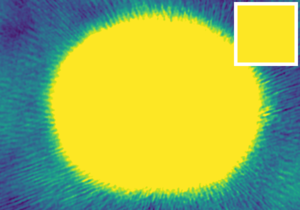}\,\hspace{5cm}}\\
 \subfloat[Multi-scale clean (residual RMS=310 $\mu$Jy/PSF)]{\includegraphics[width=5cm]{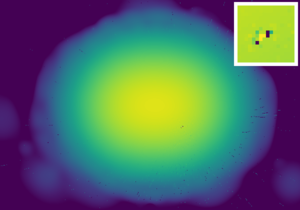}\,\includegraphics[width=5cm]{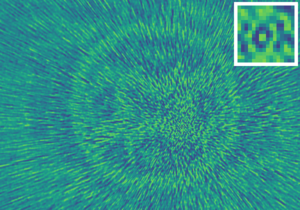}\,\hspace{5cm}}\\
 \subfloat[\textsc{moresane} (residual RMS=2000 $\mu$Jy/PSF)]{\includegraphics[width=5cm]{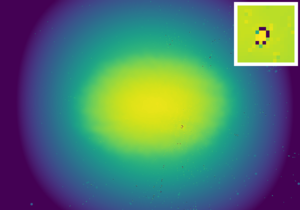}\,\includegraphics[width=5cm]{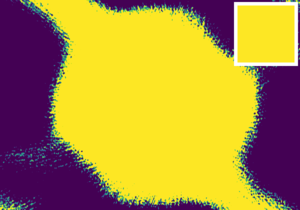}\,\hspace{5cm}}\\
 \subfloat[Multi-frequency single-scale clean (residual RMS=460 $\mu$Jy/PSF)]{\includegraphics[width=5cm]{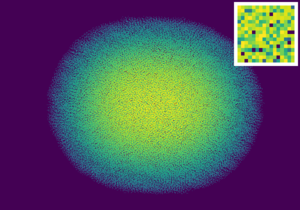}\,\includegraphics[width=5cm]{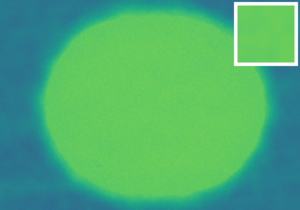}\,\includegraphics[width=5cm]{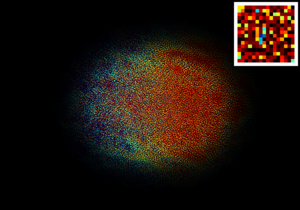}}\\
 \subfloat[Multi-frequency multi-scale clean (residual RMS=63 $\mu$Jy/PSF)]{\includegraphics[width=5cm]{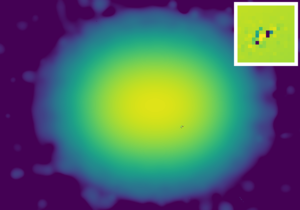}\,\includegraphics[width=5cm]{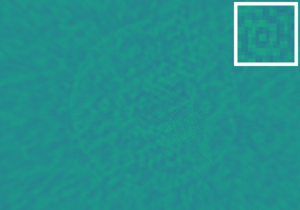}\,\includegraphics[width=5cm]{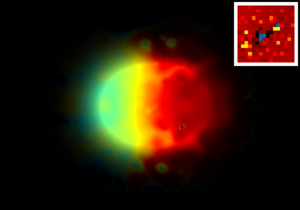}}\\\vspace*{2mm}
 \includegraphics[width=4cm]{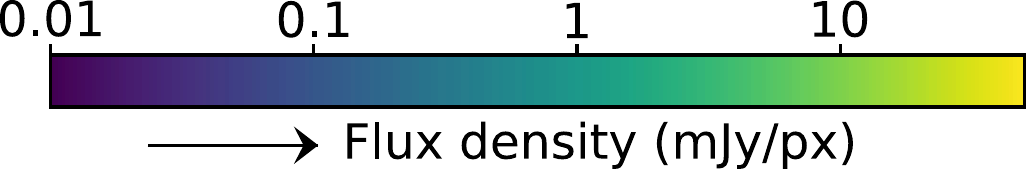}\hspace*{1cm}\,\includegraphics[width=4cm]{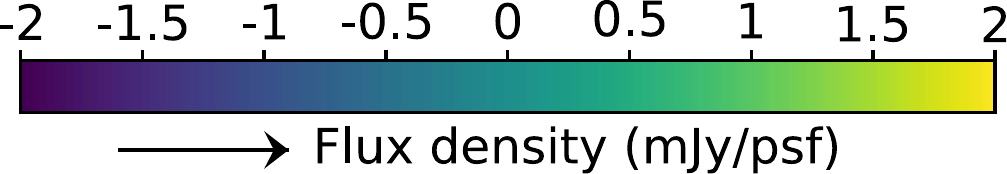}\hspace*{1cm}\,\includegraphics[width=4cm]{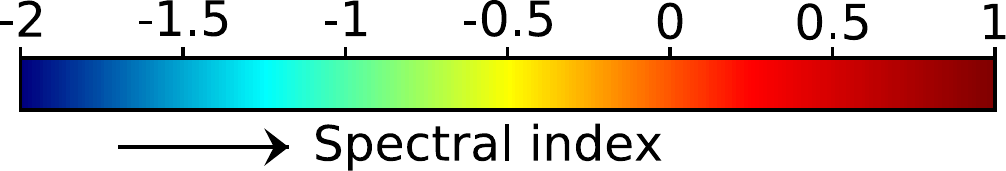}
\caption{Results of the various deconvolution methods for a test set with two overlapping Gaussians and a point source, with spectral indices of -1, 1 and -2 respectively. Left column: frequency-integrated model images; centre column: residual images; right column: modelled spectral index (only for the multi-frequency methods). Images in the same column use the same colour scale. In the spectral-index image, the colour quantifies the spectral index and the brightness quantifies the flux density. The top-right corner of each image shows a zoom-in on the point source.}
\label{fig:gaussian-simulation-results}
\end{figure*}

\begin{figure*}
\begin{center}
\subfloat[Restored image (multi-scale clean)]{\label{fig:cyga-mfs-results-restored}\includegraphics[height=4.2cm]{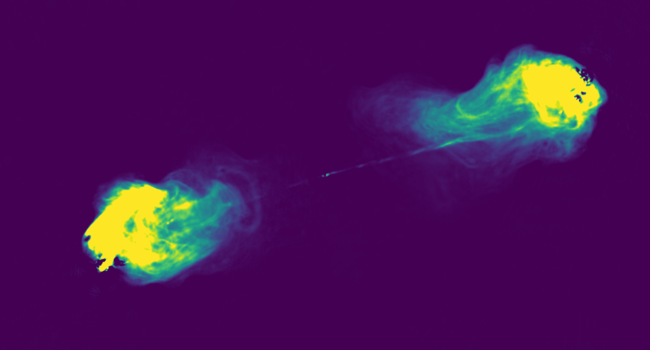}}\,
\includegraphics[height=4.2cm]{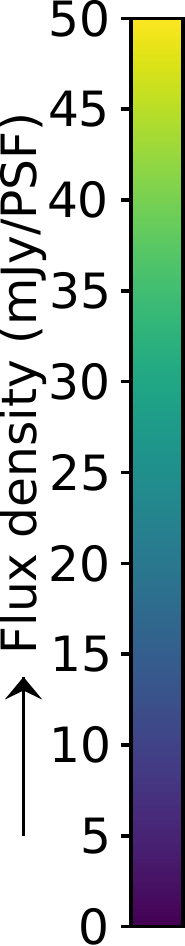} \\
\subfloat[Manually-masked H{\"o}gbom clean residual]{\includegraphics[height=4.2cm]{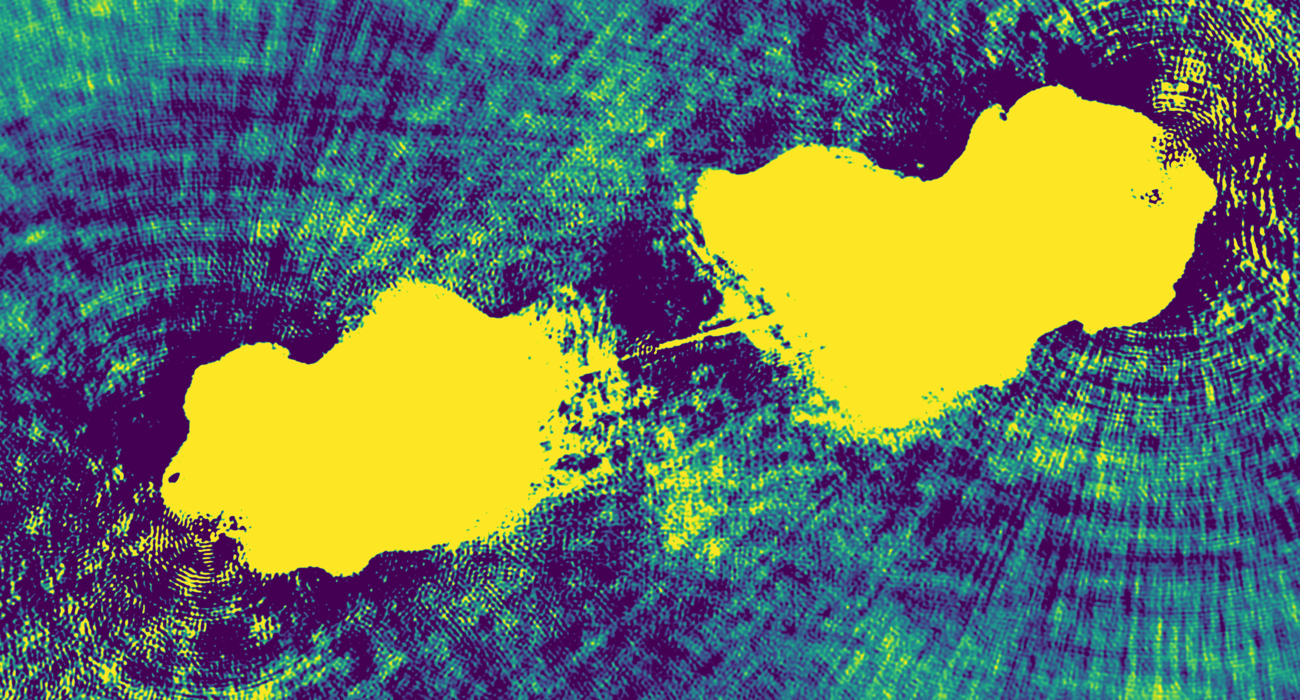}}\,
\subfloat[Residual for \textsc{moresane} with frequency interpolation]{\includegraphics[height=4.2cm]{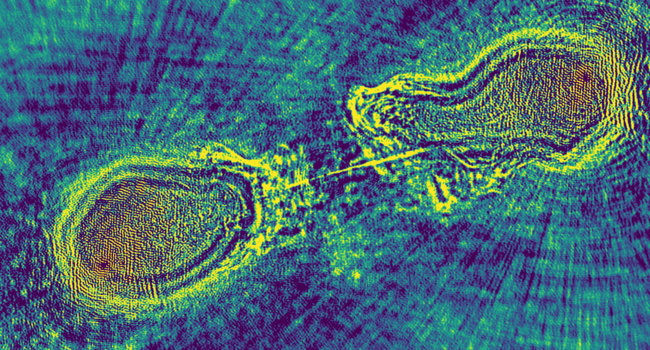}}\,
\includegraphics[height=4.2cm]{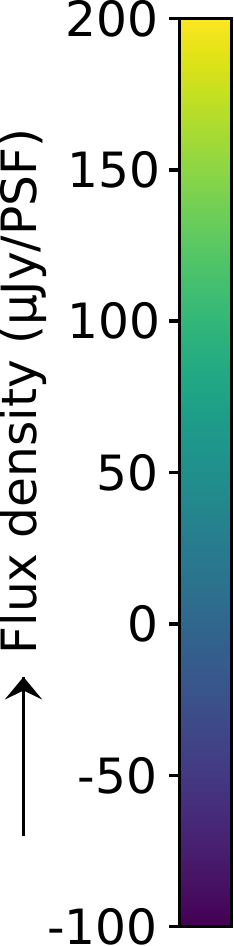}\\
\subfloat[Multi-scale clean residual]{\includegraphics[height=4.2cm]{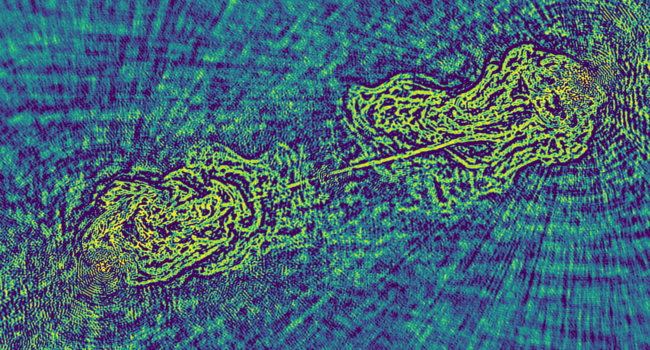}}\,
\subfloat[Multi-scale clean residual with scale-dependent mask]{\includegraphics[height=4.2cm]{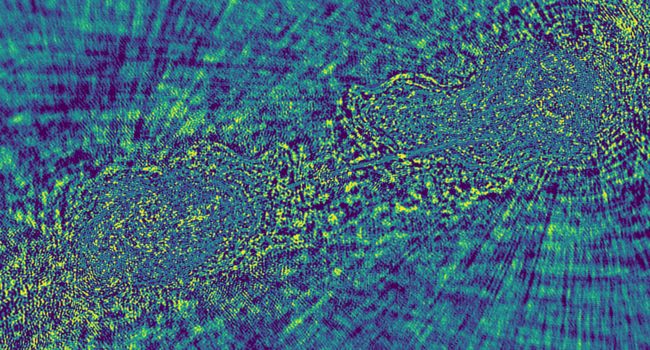}}\,
\includegraphics[height=4.2cm]{img/axes/cyga-residual-axis}
\end{center}
\caption{Comparison of achieved dynamic range for deconvolution of Cygnus A.}
\label{fig:cyga-mfs-results}
\end{figure*}

\subsection{Multi-frequency results on simulations}
To test the effectiveness of the algorithms for wide-band imaging, we use a simulation of two Gaussians with spectral indices of -1 and 1 and a point source with a spectral index of -2, similar to the simulation of \citet{rau-msmfs-2011}, although with a different array layout and smaller bandwidth. Our simulated bandwidth is 30~MHz with a central frequency of 149~MHz. The model is analytically predicted and evaluated for the MWA antenna array. The simulated synthesis time is 2 minutes, targeting zenith. The Gaussians have a size of 2$\degree$ at full-width half-maximum and are 1.5$\degree$ apart, while the point source is positioned 45' below the centre of the right Gaussian. The Gaussians has an integrated flux density of 400~Jy and peak flux of 35 mJy while the point source is 1~Jy. Deconvolution is stopped when the residual peak is lower than three times the standard deviation of the residual. Because this implies that the threshold is relative to the level of the residuals, a method that leaves fewer residuals will also perform a deeper clean.

For some positions and sizes of the simulated Gaussians, we notice that \textsc{moresane} and \textsc{iuwt} do not converge on the simulated observation, and we were not able to make a deeply deconvolved image with either method. While this simulation does not contain any calibration errors, we conclude that spectral variations can also prevent these methods from working properly. We have chosen a position and size for the Gaussians such that \textsc{moresane} still converges, but we have not been able to make \textsc{iuwt} converge. Therefore, \textsc{iuwt} is left out of the results.

Fig.~\ref{fig:gaussian-simulation-results} shows the produced residuals, model images and model spectral-index maps. Both the single-scale H\"ogbom and multi-scale clean results show lower residuals when using multi-frequency clean. The single-scale clean leaves the characteristic ``plateau'' of residual structure. Because the multi-frequency single-scale clean leaves fewer artefacts behind, it produces results that have been cleaned deeper, which lowers the residual plateau. Similarly, the multi-scale multi-frequency clean leaves less residual structure behind compared to the multi-scale single-frequency clean.

To quantify the residuals, we have measured the RMS in a box about 5$\degree$ away from the Gaussians. The measures values are given in Fig.~\ref{fig:gaussian-simulation-results}. The multi-frequency multi-scale method achieves the lowest residual noise level of 64 $\mu$Jy/PSF. Compared to a starting RMS of 280~mJy/PSF in the dirty image, the method is able to reduce the RMS by a factor of 4400. By using the automatic scale-dependent masking technique, we are able to clean slightly further and reach a noise level of 55~$\mu$Jy/PSF, or a factor of 5100 lower than the dirty image RMS.

Both spectral-index map results show that the multi-frequency methods produce a gradient going from approximately -1 to 1 from the left to the right of the image, which matches the input model. Due to the use of only delta scales, the single-scale H\"ogbom clean shows a very pixilated spectral-index map. The multi-scale clean captures the spectral variation more accurately. However, while the input model describes a smoothly varying spectral index, the multi-scale spectral indices shows patches in which the spectral index is off by a few tenths of spectral index points. This is not very surprising, since it is fundamentally hard to measure the in-band spectral index with a fractional bandwidth of 30~MHz/149~MHz. For example, in a previous study of in-band spectral indices with the MWA using 45~h and 60 MHz of data, the estimated error in measured spectral indices was 0.3 spectral index points \citep{offringa-mwa-deep-eor-survey}.

\subsection{Multi-frequency results on real data} \label{sec:vla-mf-example}
To demonstrate the effectiveness of the multi-frequency approaches on real data, we apply the various approaches to 
high-dynamic-range JVLA observations of the source Cygnus~A. The data were taken as part of a multi-band observational 
campaign on Cyg~A, and are used here with the permission of the investigators (R. Perley, priv. comm.) The 
particular subset of data used for this test consisted of a 832 MHz chunk of largely RFI-free bandwidth centred on 3.47 GHz,
with about 25h of total synthesis time split between A, B and C configurations.
Imaging is performed at a resolution of $6''$ and an image size of $6120\times 6120$ pixels.

Cyg~A is dominated by two extremely bright hotspots, and imaging dynamic range was initially dominated by radial 
artefacts associated with these. To make sure we were testing deconvolution of actual structure, and not simply 
the response to artefacts, we performed several rounds of self-calibration on the data, followed by 
direction-dependent (DD) solutions on the two hotspots, using a Gaussian component model fitted to the hotspots using PyBDSM. 
We then subtracted the hotspot models in the visibility domain, while applying DD solutions. The residual visibilities,
containing all the remaining structure in the field,  were then used as the basis of the tests in this section. For this
reason, the images shown contain some negative subtraction artefacts at the positions of the hotspots. 
The resulting images are limited by deconvolution artefacts rather than calibration artefacts, and thus provide an
 excellent real-life test for the methods discussed here.

The results are shown in Fig.~\ref{fig:cyga-mfs-results}. The results for the multi-frequency \textsc{iuwt} method have been left out, because the method does not converge for this observation. In the residuals, the single-scale H\"ogbom clean shows again a plateau of high values, and slightly higher residual off-source artefacts when compared to the other methods. \textsc{moresane} and multi-scale Clean show residual structure of very similar strength, but the structure of the two methods is very different: whereas \textsc{moresane} produces ringed residuals around the contours of the source, MF multi-scale clean leaves filamentary structure behind. When MF multi-scale is combined with the automated scale-dependent masking technique, the filamentary residuals are deconvolved, and this approach provides the best result in terms of residual structure. Despite the difference in residual structure, the off-source residual artefacts are of very similar levels between \textsc{moresane}, multi-scale and multi-scale with scale-dependent masking.

\subsection{Computational performance}

\begin{figure}
\begin{center}\includegraphics[width=8cm]{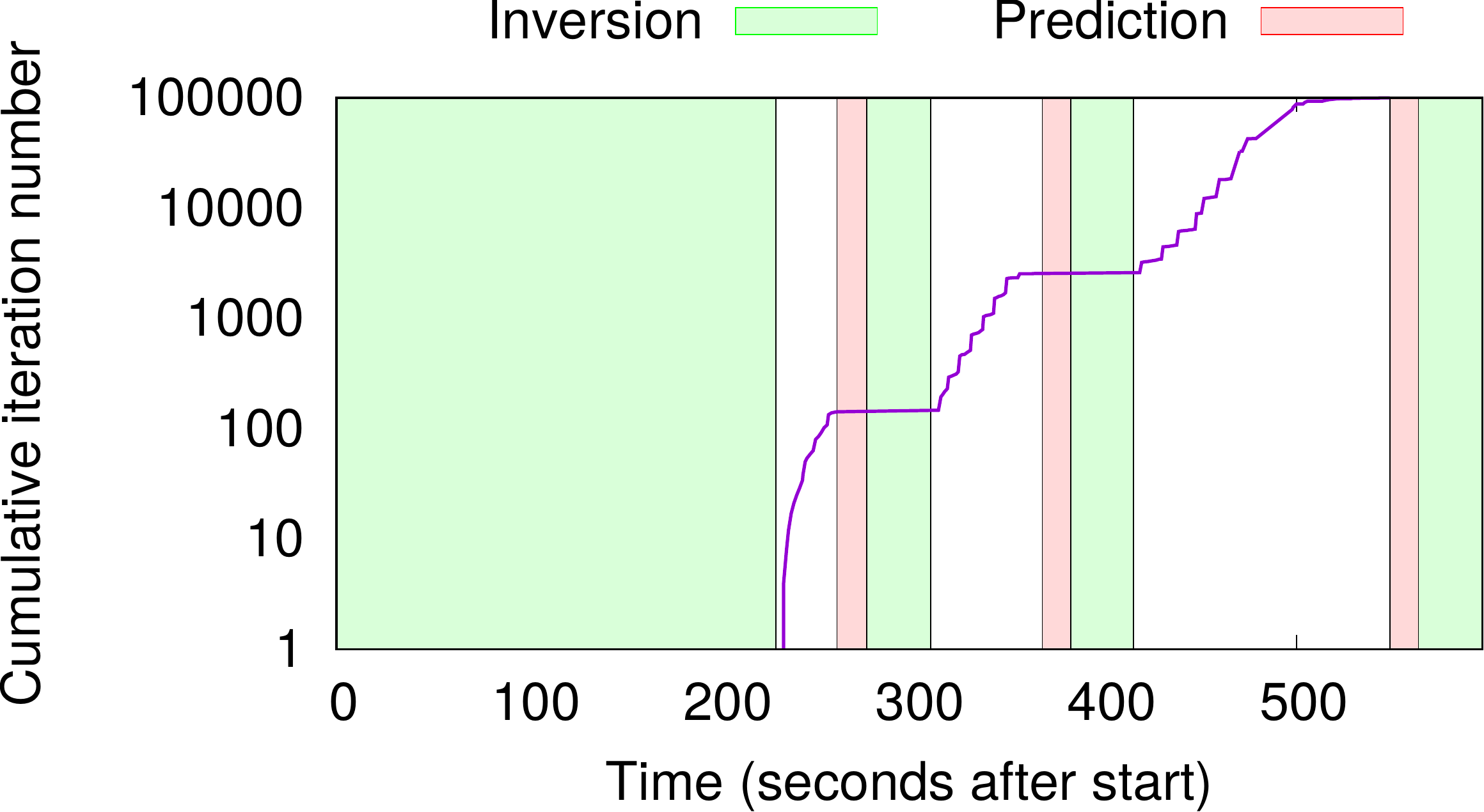}\end{center}
\caption{Example of the progression over time when using the new multi-scale clean algorithm on a $2048\times2048$ image.}
\label{fig:speedplot}
\end{figure}

In this section we analyze the computational requirements of our improved multi-scale clean algorithm by measuring the deconvolution speed.
The algorithmic improvements to multi-scale clean as discussed in \S\ref{sec:wsclean-multiscale} change the speed of the minor iterations only, and do not change the speed or number of required inversions and predictions. Therefore, we focus only on the speed of minor iterations, and ignore the inversion and prediction calculations.

We compare the cleaning speed, expressed as a number of iterations per second between the \textsc{casa} and \textsc{wsclean} multi-scale methods. To perform a comparison, it is important to take into account that the computational performance of our optimized multi-scale clean algorithm varies over time. Initially, the algorithm deconvolves only the brightest sources, which causes the maximum peak to drop quickly, and only a few iterations are performed in the subminor loop before the loop ends and the algorithm reanalyzes what scale to continue the deconvolution with. Once the brightest peaks have been removed, the subminor loop can perform more iterations before the scales have to be reanalyzed, and its speed increases. The algorithm described by \citet{multiscale-clean-cornwell-2008} as implemented in \textsc{casa} does have a constant cleaning speed.

The \textsc{casa} algorithm subtracts a scaled PSF from the residual image in each minor iteration, and for each scale. The theoretical time complexity is therefore
\begin{equation}
 \textrm{\textsc{casa} time complexity} = \mathcal{O}\left( N_\textrm{i} N_\textrm{s} N_\textrm{p} \right),
\end{equation}
with $N_\textrm{i}$ the number of iterations, $N_\textrm{s}$ the number of scales and $N_\textrm{p}$ the number of pixels. This excludes the convolutions with the scale kernel, which need to be performed once. For \textsc{wsclean}, the time complexity is given by
\begin{equation} \label{eq:wsclean-time-complexity}
 \textrm{\textsc{wsclean} time complexity} = \mathcal{O}\left( N_\textrm{i} N_\textrm{p} + N_\textrm{m} N_\textrm{s} N_\textrm{p} \log N_\textrm{p} \right),
\end{equation}
with $N_\textrm{m}$ the number of times a multi-scale subminor loop is started. The first term represents the subminor iterations, while the second term represents the convolutions required at the beginning and end of the subminor loop. Theoretically, the new algorithm lowers the per-iteration cost and replaces it with a more expensive per-subminor loop cost. Hence, this becomes beneficial when the number of iterations per subminor loop are high. \textsc{wsclean} has an extra logarithmic dependency on the number of pixels, which might make it seem that this could result in an increased cost for very large images. However, for larger images the ratio $N_\textrm{i}/N_\textrm{m}$ will increase, because images will have more pixels of comparable intensity, and on average more iterations can therefore be performed inside the subminor loop.

We measure the single-term (single-frequency) deconvolution speed in practice on a 40-core Intel Xeon E5-2660 v2 @ 2.20GHz CPU with 128 GB of memory, by deconvolving a $2048\times2048$ pixel image with multi-scale clean using 6 scales. Since a minor iteration of our new algorithm performs an equal amount of work to one in the multi-scale algorithm in \textsc{casa}, the average minor iteration speed can be used to derive and compare the total runtime.

For cleaning in \textsc{casa}, we measure a wall-clock time of 348~s over 10,000~iterations, resulting in an average speed of 29 iterations/second. This deconvolution speed is measured without including the costs involved in the inversion, prediction and the initialization of the scale kernels. The speed is constant over time. Fig.~\ref{fig:speedplot} shows the number of performed iterations as function of time for our multi-scale algorithm implemented in \textsc{wsclean}. Three major iterations are performed with 142, 2409 and 97449 minor multi-scale iterations. The average speed in these major iterations is 5, 44 and 748 iterations/second, respectively.

Comparing the two algorithms on a cleaning task of 100,000 iterations, we find that the total wall-clock time to perform 100,000 clean iterations is 223~s in \textsc{wsclean} and 3480~s in \textsc{casa}. Hence, in this imaging configuration, the speed of the optimized multi-scale algorithm is over an order of magnitude larger than Cornwell's multi-scale algorithm. 100,000 iterations is a reasonably small number for high-resolution LOFAR, MWA or VLA images. About 2~M clean iterations were required to deconvolve the high-resolution LOFAR image presented in \citet{mahony-lockman-hole-2016}, while the VLA CygA set used in \S\ref{sec:vla-mf-example} required 872~K iterations. For higher number of total iterations, the number of iterations per subminor loop will increase, hence the fractional difference between the two methods will increase further.

We continue by comparing the speed of the multi-frequency multi-scale methods of \textsc{casa} and \textsc{wsclean}. For the \textsc{casa msmfs} method \citep{rau-msmfs-2011} with \textsc{nterms=2} and otherwise equivalent settings, we measure a performance of 0.42 iterations/s, implying it takes 66~h to perform 100,000 iterations. Using our improved multi-frequency algorithm, the run-time required for 100,000 iterations is 265~s, 349~s and 600~s with 2, 4 and 8 channels respectively. The final major iteration achieves a performance of 771, 498 and 323 iterations/s. In multi-frequency mode, the number of computations scales linearly with the number of output channels. However, when more channels are joinedly deconvolved, it is possible to perform more extensive parallelization of operations such as the convolutions, which explains why the algorithmic speed does not relate to the output channels in a linear fashion.

Both \textsc{casa} and \textsc{wsclean} use multi-threading inside the clean algorithm, although both implementations are only able to use a few of the 40 cores effectively. To make it clear these test results indeed are the result of a faster algorithm and not an improvement in the parallelization of the implementation, we have repeated these tests on a 4-core laptop. Both implementations were about a factor 2--3 slower on this laptop, hence the fractional difference between the two methods is comparable and not dominated by different methods of parallelization. While differences in implementation can still account for small differences in performance, most of the improvement comes from the new algorithm.

In both algorithms, the required time for a single iteration scales approximately with the number of pixels. The required number of iterations also scales approximately with the number of pixels, meaning that the time spent in the minor loops scales approximately with the squared number of pixels. If the image were to be independently deconvolved using $N$ facets, each facet needs to perform on average $N$ times fewer iterations with $N$ times fewer pixels, thereby reducing the computations required for the entire image by a factor of $N$. This is more complicated and currently not (internally) implemented in \textsc{wsclean}, but it lowers the required computations and makes the problem easier to parallelize. If the clean performance becomes a bottleneck for very large images, e.g. SKA-sized images, faceted cleaning is a trivial further improvement. Currently the cleaning is faster than inversion and prediction in most scenarios (e.g. Fig.~\ref{fig:speedplot}), and the spent time on imaging becomes more dominated by inversion/prediction for larger image sizes and visibility data volumes. Therefore, it is less pressing to speed up cleaning further compared to improving the performance of the inversion and prediction tasks.

While the multi-scale algorithms can be compared by their minor iteration speed, the Moresane and IUWT algorithms are not structured with a similar minor loop. Therefore, to give an indication of how these algorithms compare, we measure the wall-clock time used by these algorithms to do a full deconvolution with equal imaging settings, with the same data that was used to measure the multi-scale performance. As before, we do not include the time spent on gridding the visibilities. Performing a single-frequency deconvolving of the $2048\times2048$ image with \textsc{moresane} takes 256~min, while \textsc{iuwt} takes 44~min. In multi-frequency mode with 8~channels, \textsc{iuwt} takes 179~min. These values can be compared to the 223~s required to perform a single-frequency deconvolution with our new multi-scale algorithm; and 600~s to perform a multi-frequency deconvolution. In this comparison, we have used the CPU version of \textsc{moresane}. The \textsc{pymoresane} implementation \citep{kenyon-2015-pymoresane} also provides a faster GPU implementation.

\section{Discussion \& conclusions} \label{sec:conclusions}

We have demonstrated a new multi-scale algorithm that, when comparing the speed of the minor iteration of a single-frequency deconvolution test-case, is approximately an order of magnitude faster compared to the \textsc{casa} multi-scale implementation, with a similar deconvolution quality. By using the joined-channel cleaning technique described in \S\ref{sec:joined-channel-deconvolution}, this method can be used for wide-band multi-scale deconvolution. Compared to the \textsc{msmfs} algorithm implemented in \textsc{casa}, the minor loop of the algorithm is two to three orders of magnitude faster. We have also presented a new scale-bias function with favourable properties.

By combining the multi-frequency multi-scale algorithm with the automated scale-dependent masking technique described in \S\ref{sec:automated-masking-description}, one can avoid the common residual structures that cleaning leaves behind, without making the algorithm unstable. Using the masking technique, we have been able to reach the thermal noise level in the MWA observation of the complicated Vela/PupA field. Deconvolving all flux is particularly important in self-calibration loops, which would otherwise suffer from a self-calibration bias. While scale-independent masking is a common technique in radio interferometry, scale-dependent masking limits the degrees of freedom of Cleaning, and thereby increases its stability. Furthermore, because this type of masking is automated, it removes the manual interaction and judgement of an astronomer to select significant features. By avoiding manual interaction, the process is reproducible and made more scientific.

The Clean-like algorithms are significantly faster than the more complex compressed sensing or sparse optimization algorithms, and even on enormous fields such as high-resolution LOFAR sets with multiple frequencies, the Clean algorithm is normally not the bottleneck in data processing. If necessary, the Clean family of algorithms are easy to distribute, it being trivial to distribute over the spatial dimension. Several authors have already parallelized Clean by running Clean in a faceting scheme \citep{cotton-calim-2014,vanweeren-factor-2016}.

Qualitatively, compressed sensing methods such as \textsc{moresane} result in model images that occasionally contain sharper, visually more pleasing details that are more representative of the underlying source structure. Nevertheless, due to its instability when real data is used, the \textsc{moresane} and \textsc{iuwt} convex optimization algorithms are only useful in very specific cases. So far, most compressed sensing methods for radio interferometric deconvolution have only been demonstrated on relatively simple test cases, that do not include calibration errors or $w$-terms, and with a small number of visibilities and a small image size (e.g., \citealt{carrillo-2014-purify}, \citealt{junklewitz-2016-resolve}, \citealt{vijay-2017-dimreduction}).
As was shown in this paper, the \textsc{moresane} compressed sensing approach does well in simple cases, but performs less well on data with calibration errors or with artefacts from spectral curvature. It is therefore important to start testing compressed sensing methods on real imperfect data, with realistic data volumes and image dimensions, to learn whether these methods are applicable to a more common astronomical setting.

\section*{Acknowledgements}
We would like to thank A. G. de Bruyn, F. de Gasperin and R. van Weeren for providing helpful discussions. A. R. Offringa acknowledges financial support from the European Research Council under ERC Advanced Grant LOFARCORE 339743. Research of O. Smirnov is supported by the South African Research Chairs Initiative of the Department of Science and Technology and National Research Foundation.

\DeclareRobustCommand{\TUSSEN}[3]{#3}

\bibliographystyle{mnras}
\bibliography{references}

\label{lastpage}

\end{document}